%% file: ex_article.tex
\documentclass[final,onefignum,onetabnum]{siamart190516}

\usepackage{listings}
\usepackage{graphicx}
\usepackage[section]{placeins}
\usepackage{amsmath}
\usepackage{amssymb}
\usepackage{commath}
\usepackage{subfig}
\usepackage{bm}
\usepackage{enumitem}
\usepackage{siunitx}
\usepackage{multirow}
\usepackage{booktabs,array}
\usepackage{colortbl}

\usepackage{tikz}
\usetikzlibrary{shapes,arrows,calc}

\newcommand{\R}{\mathbb{R}}

\newcolumntype{C}{>{\centering\arraybackslash} m{0.3\columnwidth} }

\input{ex_shared}

\ifpdf
\hypersetup{
  pdftitle={A Tailored Convolutional Neural Network for Nonlinear Manifold Learning of Computational Physics Data using Unstructured Spatial Discretizations},
  pdfauthor={J. Tencer and K. Potter}
}
\fi

\begin{document}

\maketitle

\begin{abstract}
We propose a nonlinear manifold learning technique based on deep convolutional autoencoders that is appropriate for model order reduction of physical systems in complex geometries.  Convolutional neural networks have proven to be highly advantageous for compressing data arising from systems demonstrating a slow-decaying Kolmogorov n-width.  However, these networks are  restricted to data on structured meshes.  Unstructured meshes are often required for performing analyses of real systems with complex geometry.  Our custom graph convolution operators based on the available differential operators for a given spatial discretization effectively extend the application space of deep convolutional autoencoders to systems with arbitrarily complex geometry that are typically discretized using unstructured meshes.  We propose sets of convolution operators based on the spatial derivative operators for the underlying spatial discretization, making the method particularly well suited to data arising from the solution of partial differential equations.  We demonstrate the method using examples from heat transfer and fluid mechanics and show better than an order of magnitude improvement in accuracy over linear methods. 
\end{abstract}



\section{Introduction}
\label{sec:Intro}

High-fidelity computational simulations have become an integral part of the engineering design process.  For many applications, such as turbine design, these simulations require solving a set of partial differential equations (PDEs), whose accurate solution often requires a fine spatial resolution that creates a large state-space dimension for the resulting system of ordinary differential equations (ODEs).  As a result, numerical integration of the resulting system of ODEs is computationally costly and, in some circumstances, (such as many-query or time-critical problems) this is infeasible.

For this reason, it is desirable to generate low-dimensional surrogate models which may be solved quickly and accurately.  These surrogates are usually either data-driven surrogates (e.g. recurrent neural networks \cite{zhang2020physics, yu2019non, jia2019physics, maulik2020reduced, nakamura2021convolutional}, dynamic mode decomposition \cite{tu2013dynamic, williams2015data, rowley2009spectral, azencot2020forecasting}) or projection-based reduced-order models (pROMs) \cite{Kerschen05themethod, Benner15asurvey}.  Data driven surrogates are typically much faster and easier to implement than pROMs while pROMs may provide improved accuracy and stability.  Regardless of type, these surrogate models involve two stages: (1) a computationally expensive offline stage that involves computing state snapshots at different times for samples of the parameter space, using those snapshots to define a low-dimensional manifold on which the solution evolves, and in the case of data-driven surrogates, learning the dynamics, and (2) a computationally inexpensive online stage, computing approximate trajectories restricted to lie on the prescribed manifold.  In this paper we focus exclusively on the first (offline) stage and particularly upon the generation of low-dimensional manifolds which approximate solution variability.

The vast majority of research in this area has focused on the use of linear solution manifolds selected via balanced truncation \cite{Serkan2004, Mehrmann2005, HEINKENSCHLOSS2011559, Benner2016}, rational interpolation \cite{Beattie2017, EbertStykel2007}, or proper orthogonal decomposition (POD) \cite{CARLBERG2013623, WILLCOX2002, Benner15asurvey, carlberg2010, Chinesta2011ASR, grle2019model}.  These types of projection methods can be expected to be very accurate with a very low-dimensional manifold for problems that exhibit fast decaying Kolmogorov $n$-width.  However, many problems (particularly advection-dominated problems) exhibit slowly decaying Kolmogorov $n$-width requiring a high-dimensional linear manifold to achieve accurate results.  High-dimensional manifolds are undesirable as the dimension of the manifold is directly related to the online computational cost of the surrogate model. 

For problems with slow-decaying Kolmogorov $n$-widths, nonlinear manifolds can reduce computational costs while maintaining or improving accuracy.  This has been demonstrated for data-driven surrogates \cite{azencot2020forecasting, gonzalez2018deep, maulik2020reduced, CARLBERG2019105, Omata2019, ahmed2019memory, otto2019linearly, puligilla2018deep, yeung2019learning, lusch2018deep, scillitoe2021instantaneous, pawar2019deep, renganathan2020machine, hasegawa2020machine} as well as pROMs \cite{sonday2010manifold, chiavazzo2013reduced, winstead2017nonlinear, chiavazzo2018manifold, Pyta2016, lee2018model, lee2019deep, kim2020fast}.  
The use of a nonlinear manifold is distinct from learning nonlinear dynamics within a linear manifold \cite{wan2018data, maulik2020time, mohan2018deep, wang2019non, fresca2020comprehensive, srinivasan2019predictions, hoang2021projection}.
Nonlinear manifolds suitable for surrogate modeling may be generated by  learning approaches such as Local Linear Embedding \cite{roweis2000nonlinear}, Diffusion Maps \cite{coifman2006diffusion, nadler2006diffusion}, Laplacian Eigenmaps \cite{belkin2002laplacian, belkin2003laplacian}, or deep convolutional autoencoders \cite{masci2011stacked, chen2017deep, holden2015learning}.  In this work, we will focus on deep convolutional autoencoders, a particular architecture of convolutional neural network (CNN) that is applicable to manifold learning tasks.

CNNs are a highly advantageous architecture for this type of nonlinear manifold learning because the number of learnable weights in the network is independent of the input feature size (full-order model state space dimension), significantly reducing the computational cost \cite{Lecun1995}.  CNNs are a powerful tool that has seen success in performing varied machine learning tasks, such as image classification \cite{Szegedy_2015_CVPR, Lawrence554195, NIPS2012_4824}, image super resolution \cite{Rasti2016, Kim_2016_CVPR, Park2018}, natural language processing \cite{KalchbrennerGB14, dos-santos-gatti-2014-deep, HuNIPS2014_5550}, and even physics \cite{Das2018, StewartE16, Farimani2017, Strofer_2019, ZHU201956, zhang2017application, Guo2016, CARLBERG2019105, lee2018model, lee2019deep, Omata2019, gonzalez2018deep, morton2019parameterconditioned, AbrasHariharan2021, warey2021prediction, liu2021supervised, morimoto2021convolutional}.  Unfortunately, traditional CNNs are not currently widely applicable for surrogate modeling of real-world physical systems, because they are restricted to structured meshes (such as a sequence, 2D array of pixels, or 3D array of voxels).
The requirement of a structured mesh makes it impossible to apply traditional CNNs to problems with arbitrarily complex geometry, such as any real engineered system, without resorting to either resampling \cite{morton2019parameterconditioned} or employing overset \cite{SHERER2005459, Nakahashi2000, MEAKIN} or immersed boundary \cite{peskin_2002, MITTAL20084825} methods.  

More recently, graph convolutional networks (GCNs) \cite{Wu2019ACS,Zhou2018GraphNN} have been developed to deal with data on unstructured graphs such as those arising from wide-ranging application domains, including social networks \cite{wu2018socialgcn},  knowledge graphs \cite{henaff2015deep}, and computer vision \cite{bruna2013spectral}.  The unstructured meshes commonly used for spatial discretization of complex physical systems are a subset of these general unstructured graphs.  The analysis of data arising from computer simulation of physics problems is distinct from most types of learning tasks typically associated with GCNs.  GCNs are often designed to capture combinatorial structures like node degrees (the number of graph edges connected to a given node) that are irrelevant to the solution of PDEs \cite{Wang_Kim_Bronstein_Solomon_2019}.  Reliance on this connectivity information has caused issues in applications like computer vision where this structural information is irrelevant to the learning task \cite{Zhou2018GraphNN}.  Unlike the computer vision application, which is almost exclusively focused on triangle surface meshes (for which the most important attribute is the shape), nodal coordinates are also of little interest when attempting to reconstruct PDE solutions.  In order to reproduce the solution manifold of a set of PDEs, the spatial distribution of the field discretized by the mesh (temperature, velocity, stress, etc.) is of primary concern.  Ideally, the representation of the solution manifold encoded into the network weights should be discretization-independent which enables remeshing without requiring retraining.

Applications of GCNs in the physical sciences have thus far been primarily focused on particle physics \cite{pmlr-v70-gilmer17a, henrion2017neural, qu2019particlenet, martinez2019pileup, qasim2019learning, shlomi2020graph}, chemistry \cite{duvenaud2015convolutional, kearnes2016molecular, li2019deepchemstable, wang2019molecule, ye2019symmetrical}, biology \cite{fout2017protein, zitnik2018modeling, feinberg2018potentialnet}, and material science \cite{xie2018crystal, chen2019graph, Karamad_2020, xie2018hierarchical, zeng2018graph, zheng2018machine}. 
In all cases, either the type of data considered or the learning task were distinct from what we propose here.
To our knowledge, this work represents the first application of GCNs to data arising from PDE solutions using unstructured spatial discretizations.

In order to generate low-dimensional surrogate models for real engineered systems whose governing equations (advection-diffusion, Navier-Stokes, Euler, etc.) preclude the existance of a sufficiently low-dimensional linear solution manifold, a manifold learning method must be employed that is (1) practical for high-dimensional state-spaces, (2) able to learn complex nonlinear manifolds, and (3) not restricted to structured meshes.  Linear methods satisfy 1 and 3 but not 2.  Fully-connected autoencoder networks satisfy 2 and 3 but not 1.  Traditional convolutional autoencoder networks satisfy 1 and 2 but not 3.  In this work, we propose using a unique GCN that is compatible with data on unstructured meshes with arbitrary connectivity allowing the technique to be used for any discretized system.  

Our proposed manifold learning method is capable of satifying all 3 requirements and is thus well-suited to this challenging learning task.  The solution manifolds generated by this technique are compatible with any of the previously mentioned surrogate modeling techniques.  We draw inspiration from prior work in the fields of graph neural networks and geometric learning and formulate a solution for our learning task that is customized to the learning of PDE solutions while fitting within the more general established framework of GCNs.  

Additionally, we seek implementations that are applicable to generic spatial discretizations, including mesh-free methods. As such, we use operators that involve \textit{discretization-independent} learned weights. We propose sets of convolution operators based on the spatial derivative operators for the underlying spatial discretization, making the method particularly well-suited to data arising from the solution of PDEs. In the 3 exemplar problems chosen, our approach consistently produces more than an order of magnitude reduction in projection error relative to an optimal linear method for low-dimensional manifolds.  In the 1 exemplar problem with a structured mesh, our approach performs comparably to a traditional CNN autoencoder network.

The paper is organized as follows. 
The convolutional autoencoder network architecture is briefly described in \cref{sec:Autoencoders}.
Our new algorithm is described in \cref{sec:GraphConv} in the context of GCNs. 
The implementation details are given in \cref{sec:Mixing}, \cref{sec:Pool}, and \cref{sec:Arch}.
The performance of our method is summarized in \cref{sec:Cost}.
The numerical experiments and corresponding results are described in \cref{sec:Exp}.
The conclusions follow in \cref{sec:Conc}.

\section{Autoencoders for Manifold Learning}
\label{sec:Autoencoders}

Autoencoders are a well-established unsupervised manifold learning technique \cite{kramer1991nonlinear, schmidhuber2015deep}.  The autoencoder seeks to learn a mapping between a high-dimensional vector $x\in\R^n$ and a low-dimensional latent representation $\tilde{x}\in\R^{d_{latent}}$ with $n\gg d_{latent}$.  The autoencoder consists of two components, the encoder $h_{enc}:\R^n \rightarrow \R^{d_{latent}}$ and the decoder $h_{dec}:\R^{d_{latent}}\rightarrow\R^n$.  A neural network is used to approximate both $h_{enc}$ and $h_{dec}$.  The two are combined to define the autoencoder
\begin{equation}
h:x\mapsto h_{dec}(h_{enc}(x))
\end{equation}
which is desired to be an approximation of the identity map.  The autoencoder network is typically trained by minimizing $\norm{x-h(x)}$ in some norm.  Once trained, the encoder and decoder networks may be separated to provide a description of a $d_{latent}$-dimensional manifold in $\R^n$ approximating the training data.

Fully-connected networks are commonly used to approximate $h_{enc}$ and $h_{dec}$.  Unfortunately, the number of learnable weights in a fully-connected network scales with $n^2$.  Additionally, fully-connected networks are tied to a particular input dimensionality and structure.  As a result, practical considerations restrict the use of fully-connected networks to relatively low-dimensional applications with fixed geometries and spatial discretizations.  Recent large turbulence simulations have $\mathcal{O}(10^{12})$ grid points \cite{ravikumar2019gpu}.  Even less ambitious undertakings commonly use state vectors with $\mathcal{O}(10^6)$ or $\mathcal{O}(10^7)$ elements rendering fully-connected networks impractical.

CNNs take advantage of local support and weight sharing to make the number of learnable weights independent of the input dimension.  As a result, CNNs have enjoyed wide success in the processing of high-dimensional data, including PDE solution data \cite{Guo2016, liu2019novel, dias2020deepcfd, ye2020flow}.  Unfortunately, weight sharing in traditional CNNs is based on an assumed structure to the dataset. Each element $i$ in the input vector is assumed to have a well defined neighborhood $\mathcal{N}(i)$ with a fixed size and a consistent and meaningful ordering of nodes $j\in\mathcal{N}(i)$ within that neighborhood).  This assumption is valid for images or data on structured meshes, but is invalid for general unstructured spatial discretizations.  GCNs provide the same weight sharing benefits of CNNs without making the same assumptions on the structure of the data.

\section{Graph Convolutions}
\label{sec:GraphConv}

There are two main types of GCNs, spectral and spatial. Spectral GCNs interpret the graph convolution operation as the application of a filter from the perspective of graph signal processing \cite{Shuman_2013}.  In contrast, spatial GCNs interpret the convolution operation through the lens of information propagation.  The spatial approach can enable greater parallelism due to the inherrently local nature of the operation.  Although our contribution may be viewed through either lense, we derive our contribution as an extension to spectral GCNs and a special case of MoNet \cite{Monti_2017_CVPR} due to the convenience of matrix notation afforded to spectral GCNs.

In the literature on GCNs, a graph is represented by $\mathcal{G} = (\mathcal{V}, \mathcal{E})$ where $\mathcal{V}$ denotes the nodeset and $| \mathcal{V} | = n$ and $\mathcal{E}$ denotes the edgeset.
We use $v_i \in \mathcal{V}$ to denote a node and $e_{ij}=(v_i,v_j)\in\mathcal{E}$ denote an edge connecting $v_i$ and $v_j$.   We also have a node feature matrix $X\in\R^{n\times c}$ defining the $c$ attributes present for each node.

We define the adjacency matrix $A \in \R^{n \times n}$, such that $A_{ij}=1$ if $e_{ij} \in \mathcal{E}$ and $A_{ij}=0$ if $e_{ij} \not\in \mathcal{E}$.  For stability reasons, it is customary to explicitly include self-loops in $A$, i.e. $A_{ii}=1$.  The adjacency matrix is normalized using the degree matrix $D\in\R^{n\times n}$, which is a diagonal matrix with $D_{ii}=\sum_j A_{ij}$.  For GCNs, the graph Laplacian is defined as
\begin{equation}
\label{eq:graphL}
L = D^{-1/2} A D^{-1/2},
\end{equation}
and the graph convolution layer is then
\begin{equation}
X^{(l+1)} = \epsilon(L X^{(l)} W),
\label{eq:GClayer}
\end{equation}
where $\epsilon$ is a nonlinear activation function, $W\in\R^{c_l\times c_{l+1}}$ is a trainable weight matrix, and $c_l$ and $c_{l+1}$ are the number of input and output features for the layer respectively.

We observe that a network whose convolutional layers are exclusively of the form \ref{eq:GClayer} is insufficiently expressive owing to the comparatively small number of attributes per node for most common PDEs relative to what is typical in, for example social science \cite{hamilton2017inductive,kipf2016semisupervised} or knowledge graphs \cite{Hamaguchi_2018}.  Additionally, the graph Laplacian has the undesirable property of directly depending upon the local mesh connectivity (through the adjacency matrix) in a way that  prevents the learned weights from being discretization-independent.

Equation \cref{eq:GClayer} can be split up into 3 parts, a filter application or mixing step, multiplication by a learned weight matrix, and the application of the nonlinear activation function.  The mixing step, which ``mixes'' information from a node's neighborhood, is the multiplication of $L$ by the input feature matrix $X^{(l)}\in\R^{n\times c_l}$.  We generalize this mixing step by allowing for an arbitrary number $m$ of mixing operations $M_k\in\R^{n\times n}$, the results of which are concatenated along the feature/channel dimension:
\begin{equation}
F= \left[ M_0 X^{(l)}, M_1 X^{(l)}, ..., M_{m-1} X^{(l)} \right] \in \R^{n\times c_l m}.
\end{equation}
The layer is completed by the subsequent multiplication by a weight matrix and application of the nonlinear activation function
\begin{equation}
X^{(l+1)}  = \epsilon(F W),
\end{equation}
with the generalized mixing step remain unchanged except that the number of weights $W\in\R^{c_lm\times c_{l+1}}$ is increased by a factor of $m$.  The use of multiple mixing operators in this way has been previously shown to be beneficial for shape classification \cite{boscaini2016learning} and image segmentation \cite{jiang2018spherical} tasks.  The standard GCN using the graph Laplacian can then be viewed as a special case, where $m=1$ and $M_0=L$.  

It is desirable to minimize the number of mixing operators $m$ for a given problem because the number of learnable parameters for a given layer is $c_l\times m\times c_{l+1}$.  However, too few mixing operators will be shown to result in reduced accuracy.  For this reason, we propose several different predefined sets of mixing operators, such that the network may be tailored to the requirements of the application.

\section{Mixing Layers}
\label{sec:Mixing}

We propose deriving mixing operators from the spatial gradient operators defined for the underlying spatial discretization.  The existence of these gradient operators is a prerequisite to any PDE solution procedure and, thus, should be readily available for any application.  Unlike $L$, these operators should not introduce undesirable dependencies of the learned weights on the local element connectivity.  

At this time, it is instructive to draw some comparisons to the traditional convolution operator from which we draw inspiration.  Traditional convolution operators may be expressed in a way that is comparable to the description used for graph convolution layers in \cref{sec:GraphConv} which is helpful for understanding the relationships between the two.  Traditional convolution operators are usually described as the application of a kernel to an input of the same dimension.  While the discussion to follow is applicable to any convolution operator, we will focus on the common case of the application of a $3\times 3$ kernal to a 2D array (e.g., an image) for illustrative purposes.

The application of any kernel may be written as multiplication by an appropriately chosen sparse matrix.  A $3\times 3$ kernel has 9 parameters, $k_{ij}$, $(i,j)\in\lbrace -1,0,1\rbrace \times\lbrace -1,0,1\rbrace$.  For a given input image $X^{(l)}\in \R^{M\times N}$, the output image from the convolution is
\begin{equation}
\label{eq:kernel_application}
X^{(l+1)}_{m,n} = \sum_{i,j} k_{ij} X^{(l)}_{m+i,n+j}.
\end{equation}
If the boundaries are treated with zero-padding, $X_{0,n} = X_{M+1,n} = X_{m,0} = X_{m,N+1} = 0$ for all $m,n$.  If we flatten $X$ into $\text{vec}(X)\in\R^{MN}$ we may write \cref{eq:kernel_application} as a sum of matrix-vector products.  We define the linear flattening operation as 
\begin{equation}
\begin{split}
\text{vec}(X) &= \sum_{i=1}^N e_i \otimes X e_i \\
&= [X_{1,1},...,X_{M,1},X_{1,2},...,X_{M,2},...,X_{1,N},...,X_{M,N}]^T,
\end{split}
\end{equation}
with $e_i$ as the $i$-th canonical basis vector for the $N$-dimensional space and $\text{vec}^{-1}_{M,N}$ as the corresponding inverse operation.
We see that
\begin{equation}
\label{eq:vectorized_kernel_application}
\text{vec}(X^{(l+1)}) = \sum_{i,j} k_{ij} P_{ij} \text{vec}(X^{(l)}),
\end{equation}
where $P_{ij}\in\R^{MN\times MN}$ is a suitably chosen permutation matrix such that \cref{eq:kernel_application} and \cref{eq:vectorized_kernel_application} are equivalent, i.e. $\text{vec}^{-1}_{M,N}(P_{ij}\text{vec}(X))_{m,n} = X_{m+i,n+j}$.

From \cref{eq:vectorized_kernel_application} we see that the traditional $3\times 3$ convolution operator is equivalent to a set of 9 mixing operators with each mixing operator being a permutation matrix corresponding to a shift of the input image.  The kernel parameters $k_{ij}$ play the role of the learned weights.  There is a one-to-one correspondence between traditional convolutional kernels and these mixing operators.  This connection has been previously exploited in the context of data-driven PDE identification for structured spatial discretizations \cite{long2018pde, long2019pde}.  

\begin{figure}[]
 \subfloat[$I$]{%
 \includegraphics[width=0.22\columnwidth]{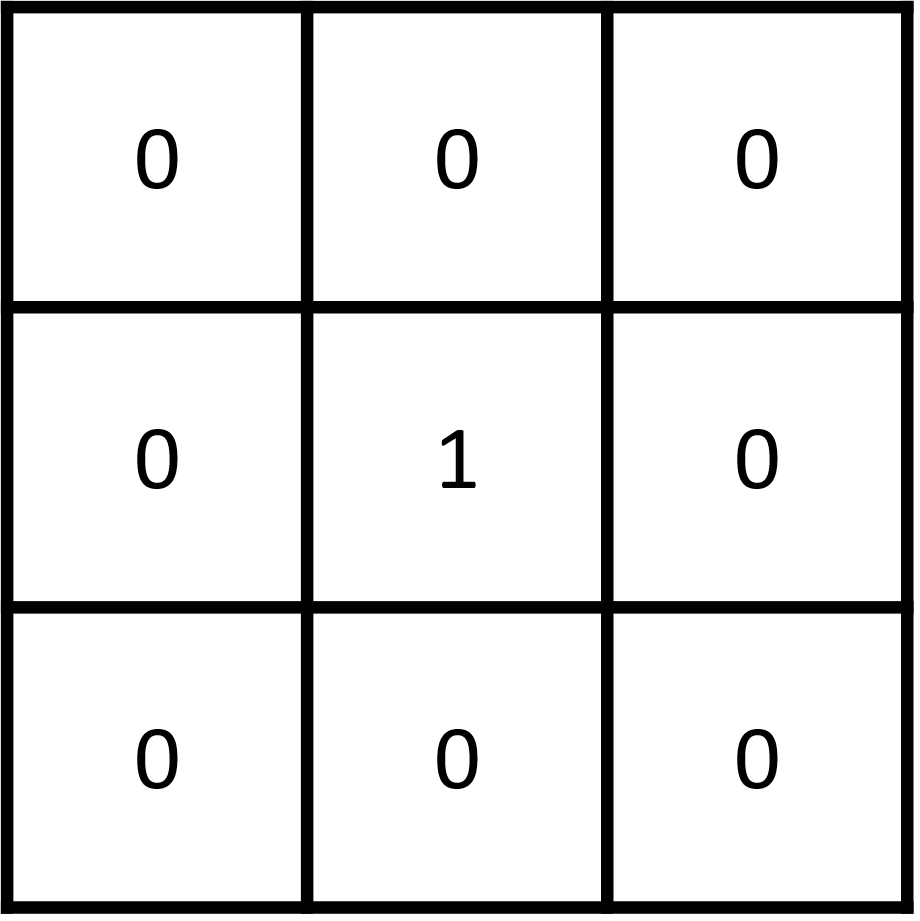}
 }
 \hfill
 \subfloat[$\nabla_x$]{%
 \includegraphics[width=0.22\columnwidth]{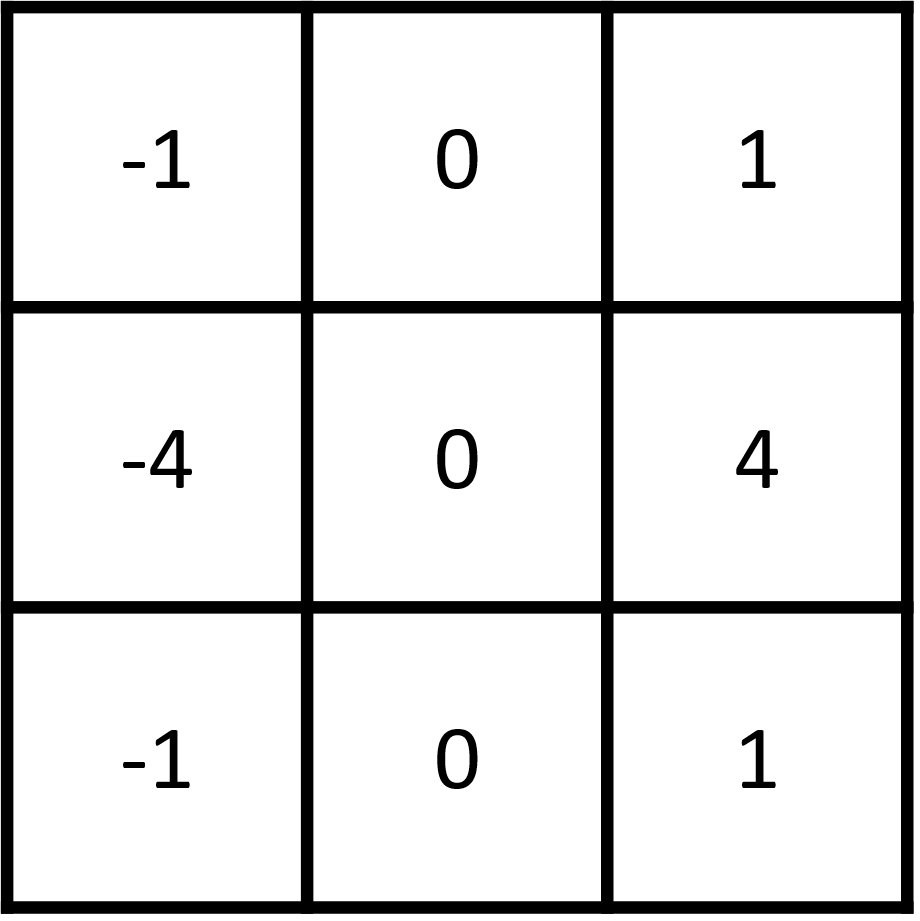}
 }
 \hfill
 \subfloat[$\nabla_y$]{%
 \includegraphics[width=0.22\columnwidth]{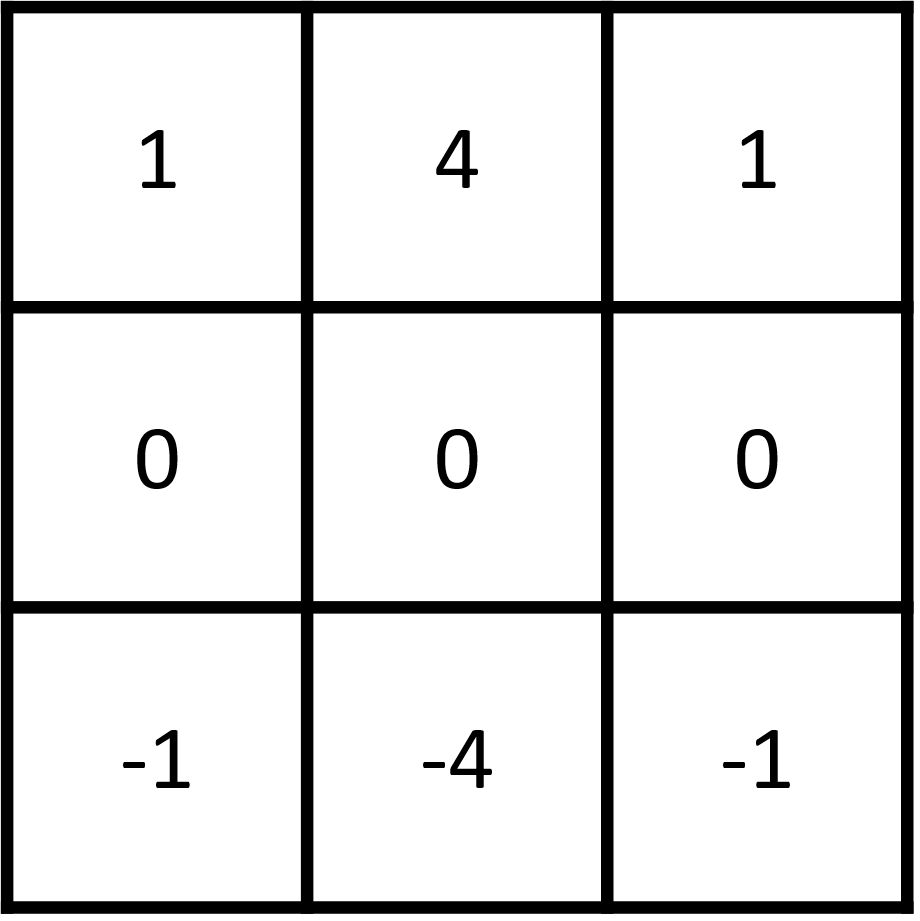}
 }
 \hfill
 \subfloat[$\Delta$]{%
 \includegraphics[width=0.22\columnwidth]{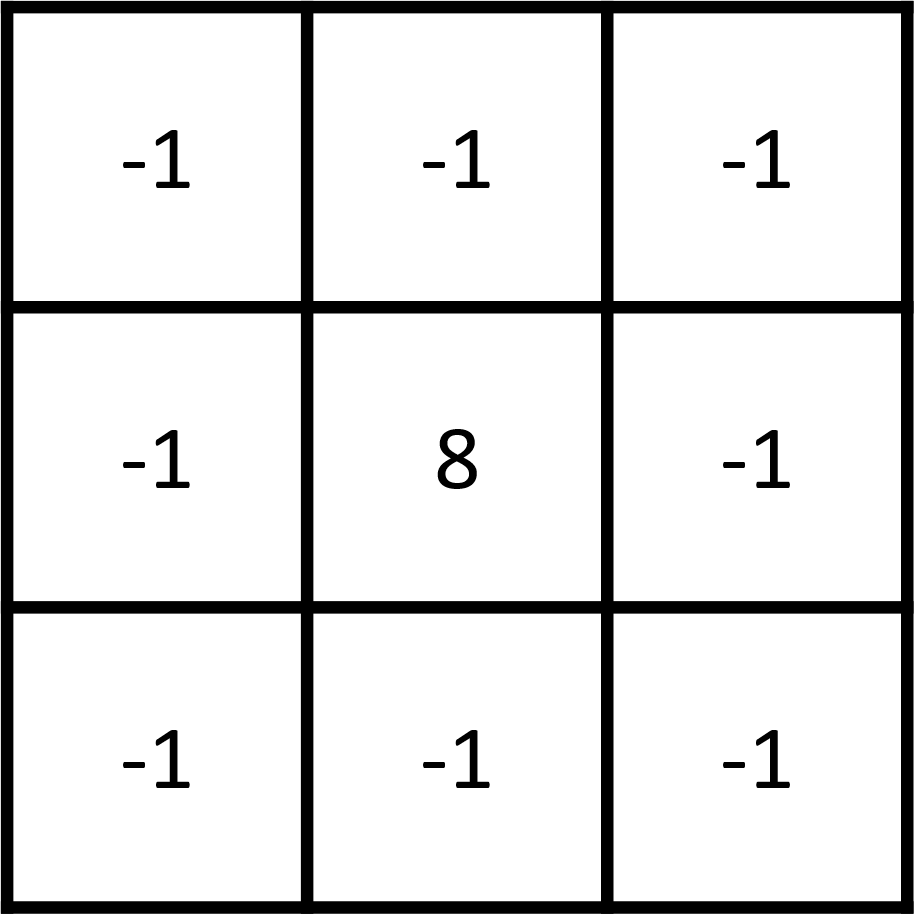}
 }
 \caption{Differential operators on a regular square mesh are closely related to common filters used for image processing.  Note that the gradient operators (b and c) are Sobel filters and that the Laplacian (c) is an edge detection filter. }
 \label{fig:kernels}
\end{figure}

If we view an $M \times N$ image as being defined on a regular rectangular mesh with nodes at the pixel centers, we can compute matrix representations of various differential operators using a linear finite element basis.  These differential operators are trivially expressed as sparse $MN \times MN$ matrices.  For interior nodes, these differential operators are equivalent to convolutions with the (unnormalized) kernels shown in \cref{fig:kernels}.
There are some differences between how these operators weight contributions at the boundary nodes compared to the normal approach of zero-padding for traditional CNNs, but these differences are not generally significant. Zero-padding can be viewed as applying a (potentially unphysical) boundary condition to the filtered outputs which has been observed to result in degraded performance near the boundary \cite{innamorati2018learning}.  The results of using finite element differential operators at the boundary is more similar to partial convolution based padding \cite{liu2018partial} although still not equivalent.  Incorporating known boundary condition information into the mixing operators is a potential area of future research.  Embedding this type of constraint has been shown to be beneficial for traditional CNNs \cite{mohan2020embedding}.

Looking at the differential operators in \cref{fig:kernels}, it quickly becomes apparent that a given set of mixing operators defines a linear subspace of the 9-dimensional space spanned by the normal $3\times3$ convolution kernel.  It has been previously observed that learned filters may be represented using a lower dimensional basis as a form of network compression \cite{qiu2018dcfnet}.  Visualizing these operators by looking at their stencils is useful for gaining intuition regarding how expressive the resulting network is likely to be.  Note, for instance, that for the particular case of a regular square grid that the Laplacian $\Delta=\nabla^2$ is a common edge-detection kernel while $\nabla_x$ and $\nabla_y$ are closely related to Sobel filters.  The $L$ operator from \ref{eq:graphL} is a box blur (if diagonal connections are included as graph edges) and may be recovered as a linear combination of the differential operators $L=I-\frac{1}{9}\Delta$.

Parameterized differential operators have been previously proposed for various learning tasks including shape classification of meshed surfaces \cite{boscaini2016learning} and image classification and segmentation on the unit sphere \cite{jiang2018spherical}.  While these previous investigations were focused on more restricted application spaces, the operators suggested therein may be more generally applicable.  In anisotropic CNNs \cite{boscaini2016learning}, a patch operator is constructed using anisotropic heat kernels.  In our framework, that roughly corresponds to mixing operators of the form
\begin{equation}
\nabla \cdot
\left(\left[\begin{array}{ccc} \delta_{0i} & 0 & 0\\ 0 & \delta_{1i} & 0\\ 0 & 0 & \delta_{2i} \end{array}\right]\nabla\right), 
i\in\lbrace 0,1,2\rbrace,
\end{equation}
where $\delta_{ij}$ is the Kronecker delta.  Parameterized differential operators have also been previously proposed for image classification and segmentation tasks on a unit sphere \cite{jiang2018spherical}.  In that work, the set of mixing layers was chosen to be $\lbrace I, \nabla_\theta, \nabla_\phi, \nabla^2=\Delta \rbrace$ with $\nabla_\theta$ and $\nabla_\phi$ corresponding to the polar and azimuthal partial derivative operators respectively.  The derivative components were chosen to align with the global spherical coordinate system.  We will present our proposed mixing operators using 2D cartesian coordinates but application to other coordinate systems and higher dimensions is trivial.  For our test problems, we have observed that using the derivative operators directly is inadvisable, due to potentially large scale differences between a function and its derivatives.  We propose normalizing the derivative operators, such that their absolute value row sums are unity:
\begin{equation}
\begin{aligned}
\widetilde{\nabla_x} &= D_x^{-1} \nabla_x       &       \quad\quad (D_x)_{ii} &= \sum_j \abs{(\nabla_x)_{ij}}   \\
\widetilde{\nabla_y} &= D_y^{-1} \nabla_y       &       \quad\quad (D_y)_{ii} &= \sum_j \abs{(\nabla_y)_{ij}}   \\
\widetilde{\Delta} &= D_\Delta^{-1} \Delta      &       \quad\quad (D_\Delta)_{ii} &= \sum_j \abs{\Delta_{ij}}  .
\end{aligned}
\label{derivatives}
\end{equation}
This type of normalization is common for GCNs.  We have also observed performance benefits in some cases by further enriching the basis.  Specifically, separating the positive and negative contributions of the gradient operators into separate operators (which is analogous to adding a learned upwinding parameter), e.g.
\begin{equation}
\begin{aligned}
(\nabla_x^+)_{ij} &= \begin{cases} 2(D_x^{-1}\nabla_x)_{ij}, & \text{for } (\nabla_x)_{ij}>0 \\ 0, & \text{otherwise} \end{cases}       \\
(\nabla_x^-)_{ij} &= \begin{cases} -2(D_x^{-1}\nabla_x)_{ij}, & \text{for } (\nabla_x)_{ij}<0 \\ 0, & \text{otherwise} \end{cases}      \\
(\nabla_y^+)_{ij} &= \begin{cases} 2(D_y^{-1}\nabla_y)_{ij}, & \text{for } (\nabla_y)_{ij}>0 \\ 0, & \text{otherwise} \end{cases}       \\
(\nabla_y^-)_{ij} &= \begin{cases} -2(D_y^{-1}\nabla_y)_{ij}, & \text{for } (\nabla_y)_{ij}<0 \\ 0, & \text{otherwise} \end{cases}      \\    
\end{aligned}
\label{eq:split_derivatives}
\end{equation}
has proven to be valuable in some circumstances.

We therefore have a large set of proposed mixing operators, a subset of which may be used for any given mixing layer.  In the examples to follow, we will refer to the following master list \cref{eq:master_list} when specifying the set of mixing operators used in a given layer.  For compactness of notation, we omit the explicit scaling of the derivative operators and state that all mixing operators are normalized, such that their absolute value row sums are unity.  The combinations of mixing operators examined in the experiments to follow will be:
\begin{equation}
\label{eq:master_list}
\begin{aligned}
    \mathcal{M}_a &= \left\lbrace I, \Delta_x, \Delta_y \right\rbrace \\
    \mathcal{M}_b &= \left\lbrace I, \nabla_x, \nabla_y, \Delta \right\rbrace\\
    \mathcal{M}_c &= \left\lbrace I, \nabla_x, \nabla_y, \Delta_x, \Delta_y \right\rbrace\\
    \mathcal{M}_d &= \left\lbrace I, \nabla_x^+, \nabla_x^-, \nabla_y^+, \nabla_y^-, \Delta \right\rbrace\\
    \mathcal{M}_e &= \left\lbrace I, \nabla_x^+, \nabla_x^-, \nabla_y^+, \nabla_y^-, \Delta_x, \Delta_y \right\rbrace\\
    \mathcal{M}_f &= \left\lbrace I \right\rbrace\\
    \mathcal{M}_g &= \left\lbrace L \right\rbrace\\
    \mathcal{M}_h &= \left\lbrace \Delta \right\rbrace\\
    \mathcal{M}_i &= \left\lbrace I, \Delta \right\rbrace.\\
\end{aligned}
\end{equation}
Note that not all possible combinations of these operators are linearly independent (ie. $\nabla_x^+-\nabla_x^-\propto\nabla_x$), so some care should be taken when selecting operators for a mixing layer to avoid duplication.  Additionally, note that $L$ is the graph Laplacian discussed in \cref{sec:GraphConv} such that $\mathcal{M}_g$ represents a traditional GCN.

\section{Pooling and Unpooling Layers}
\label{sec:Pool}

The other crucial component to CNN architectures is the pooling operation.  
Pooling reduces the spatial resolution of the information as it propagates deeper into the network.  The most common types of pooling operations in deep learning are max pooling and average pooling which take the maximum or average value of a feature in a given neighborhood.  
There have been numerous pooling operations proposed for GCNs \cite{ranjan2020asap, diehl2019edge, lee2019selfattention, cangea2018sparse, dhillon2007weighted, simonovsky2017dynamic, bianchi2020spectral}.  In this work, we will use a sort of weighted average pooling derived from bilinear interpolation.

First, we define the unpooling operation.  Given two meshes representing a coarse and a fine discretization of the same spatial domain, an $n_{fine} \times n_{coarse}$ linear interpolation operator may be defined that maps from the coarse mesh to the fine mesh.  Each row in this matrix corresponds to a node in the fine mesh.  The values in that row correspond to the weights of each node in the coarse mesh when interpolating a function defined on the coarse mesh to the spatial location of the node on the fine mesh.  This mapping from coarse mesh to fine mesh is the unpooling operator that will be used in the numerical experiments to follow.  The corresponding pooling operator is simply the transpose of the unpooling operator defined in this way.  Both of these operators are sparse matrices.

\section{Network Architectures}
\label{sec:Arch}

\begin{figure}[h]
 \subfloat[]{%
 \includegraphics[width=0.3\columnwidth]{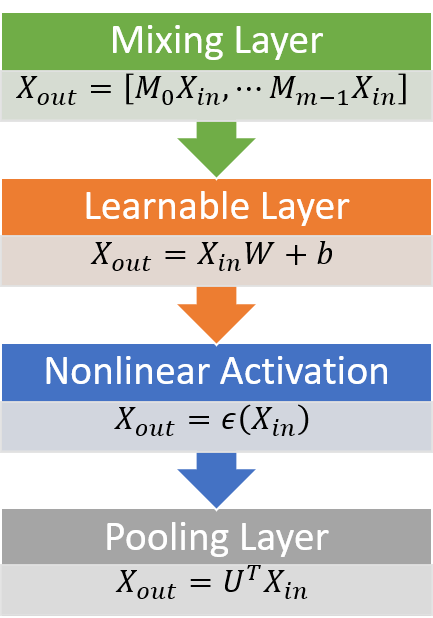}
 \label{fig:encoder_ex3}
 }
 \hfill
 \subfloat[]{%
 \includegraphics[width=0.3\columnwidth]{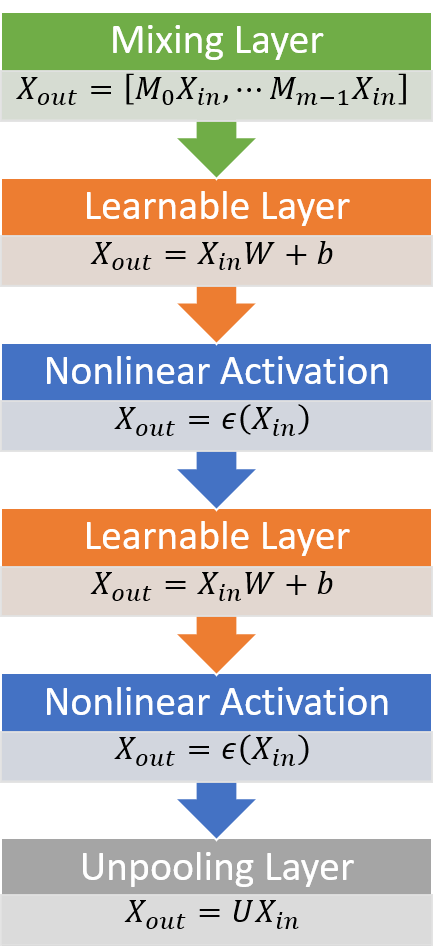}
 \label{fig:decoder_ex3_b}
 }
 \hfill
 \subfloat[]{%
 \includegraphics[width=0.3\columnwidth]{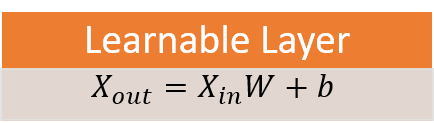}
 \label{fig:decoder_ex3_c}
 }
 \caption{The convolutional autoencoder networks are comprised of 3 fundamental components: the (a) encoder convolutional module, (b) decoder convolutional module, and (c) decoder output module.}
 \label{fig:decoder_ex3}
\end{figure}

Very similar achitectures are used for all 3 exemplar problems.  The only difference was the depth of the network. For each example, we precompute a sequence of coarsened meshes along with corresponding differential and interpolation operators. The network components are then applied at the different mesh resolutions, using the corresponding operators.  The fundamental components of the autoencoder networks are shown in \cref{fig:decoder_ex3}.  A diagram of the full autoencoder architecture is included in \cref{sec:AppendixB}.  We have observed that the encoder task is significantly less complex than the decoder task for our example applications, which is intuitive, as the decoder task is analagous to solving the system of PDEs whereas the encoder task is analgous to performing a regression to find an unknown parameter from data.  As such, we have chosen an asymmetric architecture that devotes a larger number of learned parameters to the decoder task.

The input is fed into the network as a 3D tensor of size [$B$,$C$,$n_0$], where $B$ is the batch size (snapshots per batch), $C$ is the number of input channels (number of nodal fields), and $n_0$ is the number of nodes in the input mesh.  The ordering of the nodes is unimportant as long as it's consistent across snapshots and with the row-ordering of the mixing and interpolation operators.  Scaling may be optionally applied independently for each input channel as a preprocessing step but was not required for the examples to follow.  For each network, a sequence of $d+1$ progressively coarser meshes is generated with $n_0, n_1, \cdots, n_d$ nodes.  The encoder network consists of one encoder convolutional module (\cref{fig:encoder_ex3}) for each mesh resolution, a fully connected layer to reduce the size to 64, a nonlinear activation, and another fully connected layer to reduce to the desired latent space dimension $d_{latent}$, which is treated as a hyper parameter.  All nonlinear activation layers use the ELU activation function
\begin{equation}
\epsilon(x)=max(0,x)+min(0,\alpha(exp(x)-1)),
\end{equation}
with $\alpha=1$, consistent with previously published results \cite{lee2018model}. During experimentation, the choice of nonlinear activation function was not observed to significantly impact the quality of results. For the pooling operation, we use the renormalized transpose of the interpolation operator that maps from the coarse mesh to the fine mesh as described in \cref{sec:Pool}. We will use this same interpolation operator for the unpooling layers in the decoder.

The decoder network begins with 4 fully connected layers.  The number of neurons in each layer is $64$, $4n_d$, $16n_d$, and $64n_d$, where $n_d$ is the number of nodes in the coarsest mesh.  Each fully connected layer is followed by an ELU nonlinear activation layer.  These are followed by one decoder convolutional module (\cref{fig:decoder_ex3_b}) at each mesh resolution save the final (finest) mesh.  An output layer (\cref{fig:decoder_ex3_c}) is used rather than a convolutional layer for the finest mesh.  Since no nonlinear activation functions are used in the output layer, the features may be optionally extracted at this point and treated as a nonlinear mode decomposition \cite{murata2019}, which is useful for visualization and interpretability.  While, in theory, each mixing layer could be comprised of a unique set of mixing operators; in practice, this is cumbersome to manage; so, for each experiment, a common set of mixing operators will be used for all mixing layers.

\section{Computational Cost}
\label{sec:Cost}

The computational cost may be split into two parts, the offline (training) cost and the online (inference) cost.  In terms of offline cost, deep autoencoder methods are orders of magnitude more expensive than linear methods.  In terms of online cost, the linear methods consist of a single dense matrix-vector product which requires $2d_{latent}n_0$ flops.  There have been several methods created to reduce or eliminate the $n_0$ scaling of the cost via hyper-reduction \cite{ryckelynck2005priori, carlberg2010, CARLBERG2013623, HERNANDEZ2017687}. For CNNs, each convolutional layer may be expressed as a series of sparse matrix   vector products.  Neglecting activation functions, the computational cost of a convolution at layer $l$ is approximately
\begin{equation}
\label{eq:convcost}
2 m n_l c_l c_{l+1}.
\end{equation}
Recall from \cref{sec:GraphConv} that $m$ is the number of mixing operators (9 for a traditional $3 \times 3$ convolution, $n_l$ is the number of nodes present at layer $l$, and $c_l$ and $c_{l+1}$ are the number of input and output channels respectively.  There has been a great deal of research in recent years into reducing the online cost of evaluating deep neural networks, much of it focused on network compression and pruning \cite{han2015deep, li2016pruning, anwar2017structured, zhao2019variational}.
The pooling and unpooling operations are simple sparse matrix-vector products at a cost of 2 flops per nonzero element of the interpolation matrix.

Note that the online cost only requires evaluating the decoder half of the autoencoder network and that the overall cost will depend on the network architecture (the depth, number of channels at each layer, etc.) and that \cref{eq:convcost} is valid for both traditional CNNs and the novel convolutional layer developed here.  
One could easily assume therefore that because $m<9$ for all of the mixing operator sets included in \cref{eq:master_list} that the online computational costs of our networks should be lower than those of analogous traditional CNNs.  However, the implementations of traditional convolutional layers in popular deep learning frameworks such as PyTorch and TensorFlow are highly optimized.  In contrast, support for general sparse linear algebra in these frameworks is relatively recent.  As a result, despite the reduction in flops, our implementation is noticeably slower (approximately 7x for a simple benchmark) in terms of wall clock time.  We expect that as native support in deep learning frameworks improves that the speed of GCNs will eventually match and exceed that of traditional CNNs.

\section{Numerical Experiments}
\label{sec:Exp}

We will be considering three different datasets from a range of computational physics applications.  We will be constructing autoencoders and assessing performance in terms of the degree to which the solution lies within the generated manifold.  Specifically, we will use the mean-squared error (MSE) and relative $L^2$ error between the solution snapshots and their reconstructions as our primary metrics.  For the experiments to follow, we will vary the sets of mixing operators $\mathcal{M}$ used within the network as well as $d_{latent}$.  The three datasets are:
\begin{itemize}
\item Transient advection-diffusion,
\item Unsteady laminar fluid flow past a cylinder, and
\item Inviscid supersonic flow over a wedge.
\end{itemize}

The MSE loss function is used for both training and evaluation.  This loss function weights the error at each node equally.  For regular meshes, this will correspond closely with the $L^2$ error in the reconstruction.  For highly irregular meshes, it may be advisable to use the $L^2$ or $L^1$ error directly to avoid over-emphasizing densely meshed regions.  In practice, densely meshed regions often correspond to regions of particular interest; therefore, weighting those regions more heavily in the loss calculation may be appropriate in some cases.  For the experiments to follow, MSE and $L^2$ yield similar trends and the same conclusions.  For ease of interpretation, the $L^2$ error is used for the experiments with irregular meshes.

The basis of comparison for the nonlinear manifolds will be linear manifolds generated via POD, which is optimal in the MSE metric.  In all cases the error cited for POD is the minimum projection error of the test data onto the POD basis.  For the autoencoder results, the reconstruction error used is $\norm{x-h_{dec}(h_{enc}(x))}$ which is bounded from below by the optimal projection error $\min_{\hat{x}}\norm{x - h_{dec}(\hat{x})}$.  In this way, the reported results are conservative.  

All models were trained using the AdamW optimizer \cite{loshchilov2019decoupled}  implemented in PyTorch with an initial learning rate of $5\times 10^{-3}$.  The learning rate was reduced by a factor of 0.8 if no improvement in the training loss after 10 epochs until a minimum learning rate of $10^{-5}$.  Training was terminated after 20 consecutive epochs without improvement in the training loss once the learning rate attained this minimum value.  There was no limit on the maximum number of epochs.  This termination criteria was chosen to avoid penalizing slow-converging networks.  The network weights were initialized using PyTorch defaults.  Mini-batching was used for all experiments with batch sizes of 125, 465, and 41 respectively.
The batch sizes for each experiment were chosen based on hardware limitations rather than optimized for accuracy or performance.  

In the following subsections, we will summarize the results of each experiment.  Referring back to the master list of proposed mixing operators \cref{eq:master_list}, we use all of the proposed mixing operator sets for all experiments.  Detailed results are tabulated in  \cref{sec:AppendixA}.  The data for each example was generated using a finite element spatial discretization.  The datasets used for training and testing for each experiment are included as supplemental material.  

\subsection{Transient Advection-Diffusion}

For our first example, consider the transient, 2D advection-diffusion equation:
\begin{equation}
\begin{split}
\frac{\partial\alpha}{\partial t}\left(\vec{r},t\right) =  \nabla\cdot\left(D\nabla\alpha(\vec{r},t)\right) -  \vec{u}\cdot\nabla\alpha(\vec{r},t) \\
\vec{r}\in \Omega \subset\R^2 , t\in [0,1] \\ \\
\alpha\left(\vec{r},0\right)=0 \text{ in } \Omega \\
\alpha\left(\vec{r},t\right)=1 \text{ for } \vec{r}\in\partial\Omega. 
\end{split}
\label{eq:adv-diff}
\end{equation}
The diffusivity $D$ is chosen to be $0.1$ and the advection velocity is chosen to be a unit vector, $\norm{\vec{u}} = 1$.  The equation is solved on the unit square $\Omega=(0,1)\times(0,1)$.  The direction $\theta$ of the advection velocity is varied so $\vec{u}=(\cos(\theta),\sin(\theta))$.  Training data is generated every \ang{30} beginning with the positive $x$-axis ($\theta=0$).  Separate test data is generated for angles of $\theta\in\lbrace\ang{45}, \ang{135}, \ang{225}, \ang{315}\rbrace$.  For each value of $\theta$, data is collected for $t\in(0,1]$.  Adaptive timestepping is used resulting in $\mathcal{O}(10^2)$ snapshots per value of $\theta$. 
The 5 meshes used are regular square meshes of $160\times 160$, $80\times 80$, $40\times 40$, $20\times 20$, and $10\times 10$ elements apiece. 

\begin{figure}[]
 \subfloat[Input]{%
 \includegraphics[width=0.49\columnwidth]{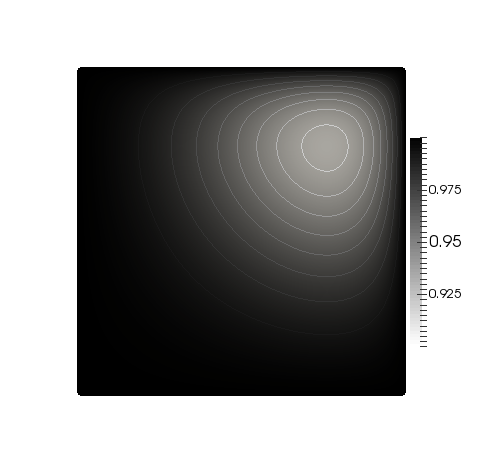}
 }
 \hfill
 \subfloat[Autoencoder ($\mathcal{M}_e$)]{%
 \includegraphics[width=0.49\columnwidth]{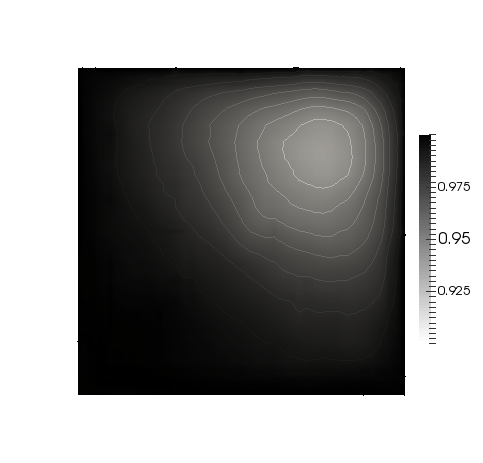}
 }
 \hfill
 \subfloat[POD]{%
 \includegraphics[width=0.49\columnwidth]{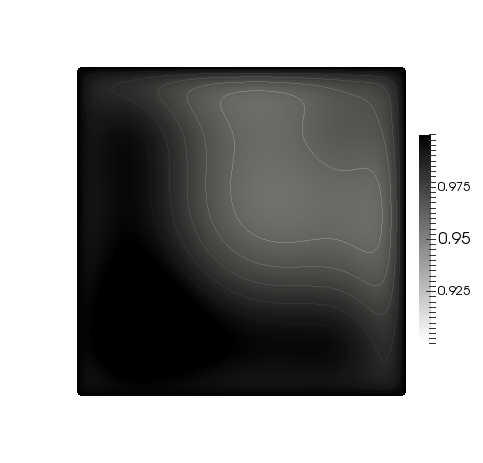}
 }
 \hfill
 \subfloat[Autoencoder (traditional CNN)]{%
 \includegraphics[width=0.49\columnwidth]{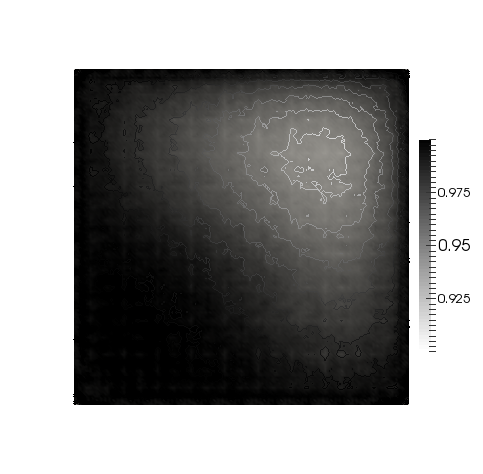}
 }
 \caption{Our graph-based autoencoder reproduces the spatial distribution of $\alpha$ much more accurately than the optimal linear (POD) manifold approach and produces qualitatively better reconstructions than even a traditional CNN autoencoder.  All 4 subfigures show the distribution of $\alpha$ at the final simulated time ($t=1$) for $\theta=45\deg$.  All reconstructions are performed with  $d_{latent}=4$.}
 \label{fig:ex1_result}
\end{figure}
 
The full results for all combinations of mixing operators, $\mathcal{M}\in\lbrace\mathcal{M}_a,\cdots,\mathcal{M}_i\rbrace$ with $d_{latent}\in\lbrace 4,16,64\rbrace$ are included in \cref{tab:training_advdiff} in  \cref{sec:AppendixA}.  There are also selected results for $d_{latent}=2$.  The results of varying $d_{latent}$ are uneven.  In general, increasing $d_{latent}$ should result in a more accurate model.  However, since the actual solution manifold for this case is fully contained in $\mathbb{R}^2$, it is perhaps expected that increasing $d_{latent}$ beyond 4 fails to continue improving the solution quality.  While it is possible to obtain accurate results for $d_{latent}=2$ for some test problems, the networks suffer in terms of generalizability; a low training loss does not imply low test loss across the board.  For POD, the solution continues to improve as expected as $d_{latent}$ is increased.  POD accuracy is comparable to the autoencoder results at $d_{latent} = 16$ and nearing machine precision for $d_{latent} = 64$.  

When constructing surrogate models, it is desirable to minimize $d_{latent}$.  For both pROMs and data-driven surrogates, reducing $d_{latent}$ results in a reduction in online computational cost.  For pROMs, the online cost is closely related to the cost of solving a least-squares problem (projecting the dynamics onto the solution manifold).  The solution is usually generated via an iterative procedure, the cost of which will be problem dependent.  If the quadratic Newton-Raphson method is used, the cost of each iteration will scale as $\mathcal{O}(d_{latent}^2)$ \cite{Erdogdu2016}.  For data-driven surrogates, the computational cost reduction will depend on the surrogate used.  

Because the mesh used for this example is a regular grid, a traditional CNN is also used.  The architecture is as similar as possible to the autoencoder architecture described in \cref{sec:Arch}.  The same number of convolutional layers with the same number of filters are used and the fully connected layers are identical.  The accuracy for both the traditional CNN and proposed extension are comparable indicating that our proposed technique effectively replicates the accuracy of the traditional CNN without relying on any underlying structure in the mesh.  Notably, the traditional CNN reconstruction suffers from some high frequency artifacts arising from the upsampling procedure, as shown in \cref{fig:ex1_result}.  It should be noted that these same high-frequency artifacts have been observed in other works using deep convolutional autoencoders \cite{aitken2017checkerboard, lee2018model, hong2020kick}.  

The results are more consistently impacted by the choice of mixing operators.  For some sets of mixing operators (especially those with $m<3$), the accuracy of the resulting model appears to be highly dependent on the random initialization of the network.  This indicates an insufficiently expressive architecture.  This is undesirable and a strong argument for preferring larger sets of mixing operators.  Mixing operator set $\mathcal{M}_e$ with $m=7$ is a consistently well-performing option.  Mixing operator sets with $m<3$ consistently perform worse than those with $m\geq3$.  Operator sets $\mathcal{M}_c$ and $\mathcal{M}_e$ are consistently the top performers, which indicates that value is added by including the anisotropic Laplace operators in the mixing operator set for this problem.  However, differences between operator sets with $m\geq3$ were less significant than the differences between those sets and sets with $m<3$, including the traditional GCN, $\mathcal{M}_g$.

\cref{fig:ex1_result} compares the reconstructions achieved via convolutional autoencoder (with mixing operator set $\mathcal{M}_e$), a linear model (POD), and a traditional CNN with a similar autoencoder architecture.  The autoencoder-based approaches consistently outperform POD.

\subsection{Unsteady Laminar Fluid Flow Past a Cylinder}

\begin{figure}[]
 \subfloat[Geometry]{%
 \includegraphics[width=0.49\columnwidth]{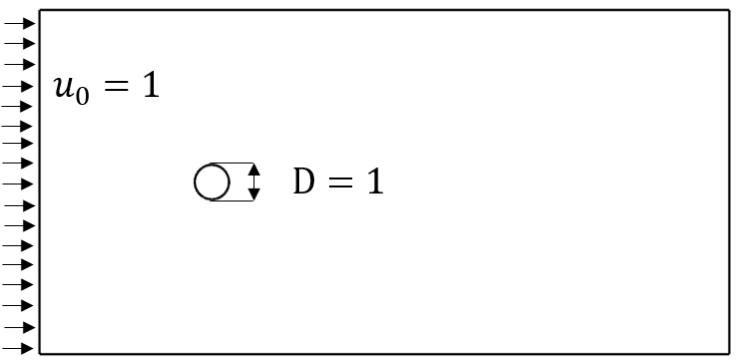}
 }
  \hfill
 \subfloat[Mesh]{%
 \label{fig:ex2_mesh}
 \includegraphics[width=0.46\columnwidth]{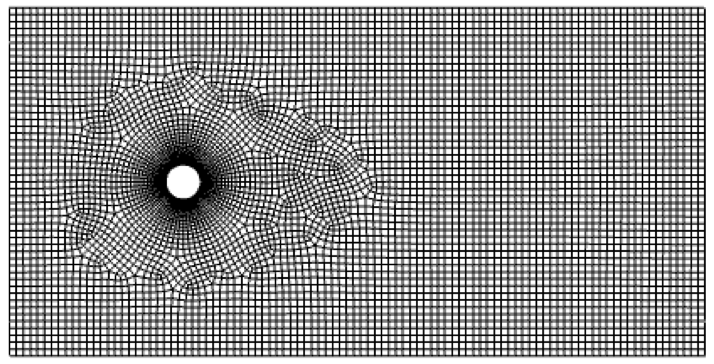}
 }
  \hfill
 \subfloat[$u_x$]{%
 \includegraphics[width=0.49\columnwidth]{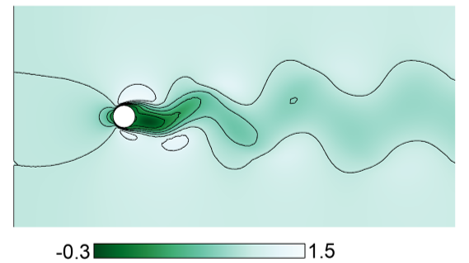}
 }
 \hfill
 \subfloat[$u_y$]{%
 \includegraphics[width=0.49\columnwidth]{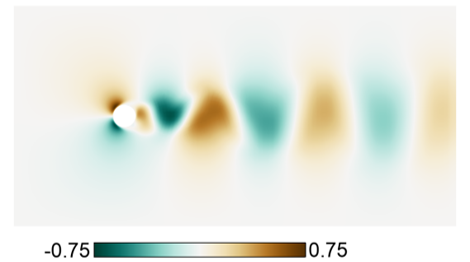}
 }
 \caption{Unsteady laminar flow past a cylinder (a) has solutions computed on an unstructured mesh (b) which exhibit the familar vortex shedding patterns visible in the (c) horizontal and (d) vertical velocities.}
 \label{fig:ex2}
\end{figure}

\begin{figure}[]
 \subfloat[Lift, $d_{latent}=4$]{%
 \includegraphics[width=0.49\columnwidth]{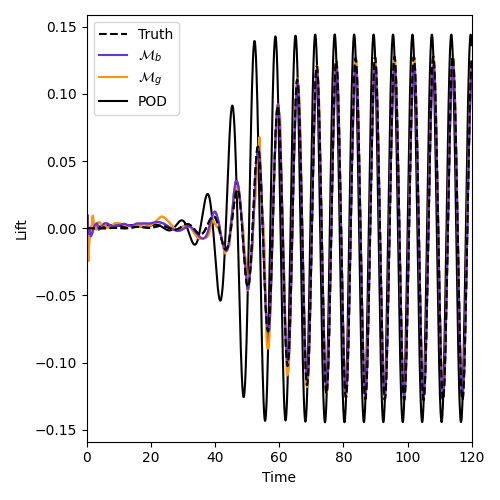}
 }
 \hfill
 \subfloat[Relative $L^2$ error in velocity]{%
 \includegraphics[width=0.49\columnwidth]{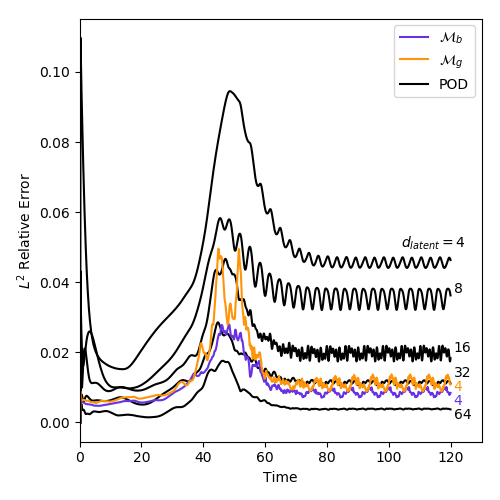}
 \label{fig:ex2_b}
 }
 \caption{(a) Autoencoder reconstructions with $d_{latent}=4$ are significantly more accurate than POD in terms of (a) transient lift generated from the reconstructed pressure and (b) transient $L^2$ error in the velocity distribution for $Re=150$.  The autoencoder accuracy with $d_{latent}=4$ is comparable to POD with $d_{latent}=32$.  Additionally, the differential operator approach ($\mathcal{M}_b$) clearly outperforms the traditional GCN ($\mathcal{M}_g$). }
 \label{fig:ex2_lift}
\end{figure}

The second example involves unsteady laminar flow past a cylinder that generates the familiar vortex shedding pattern known as a von K\'arm\'an vortex street.  The governing equations for this system are the incompressible Navier-Stokes equations:
\begin{equation}
\begin{aligned}
\frac{\partial \vec{u}}{\partial t} + \left(\vec{u}\cdot\nabla\right)\vec{u} - \nu\nabla^2\vec{u}&= -\frac{1}{\rho}\nabla p \\
\nabla\cdot\vec{u}=0.
\end{aligned}
\end{equation}
The geometry is that of a cylinder in cross-flow, as shown in \cref{fig:ex2}.  The cylinder has a diameter of $1$ and the free stream velocity is $1$.  The kinematic viscosity $\nu$ is varied such that the Reynolds number is between $100$ and $400$.  Symmetry boundary conditions are applied at the top and bottom edges of the domain and an open pressure boundary condition is applied at the outlet.  Solutions are generated on the unstructured mesh of 6384 quad elements shown in \cref{fig:ex2_mesh}.  Simulations are run with adaptive time stepping with a termination time of $120$ resulting in $\mathcal{O}(10^3)$ solution snapshots generated per Reynolds number value.  Training data is generated for Reynolds numbers of $100$, $200$, $300$, and $400$.  Separate test data is generated at Reynolds numbers of $150$, $250$, and $350$.  The fields considered for reconstruction are the horizontal velocity $u_x\in(-0.4,1.4)$, vertical velocity $u_y\in(-0.75,0.75)$, and pressure $p\in(-0.75,0.6)$.

\begin{figure}[h]
 \subfloat[Re=150]{%
 \includegraphics[width=0.49\columnwidth]{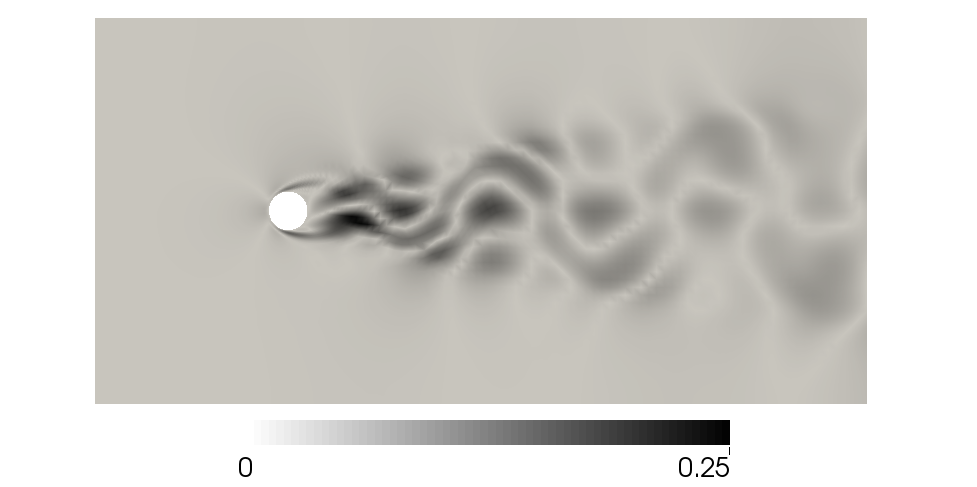}
 \hfill
 \includegraphics[width=0.49\columnwidth]{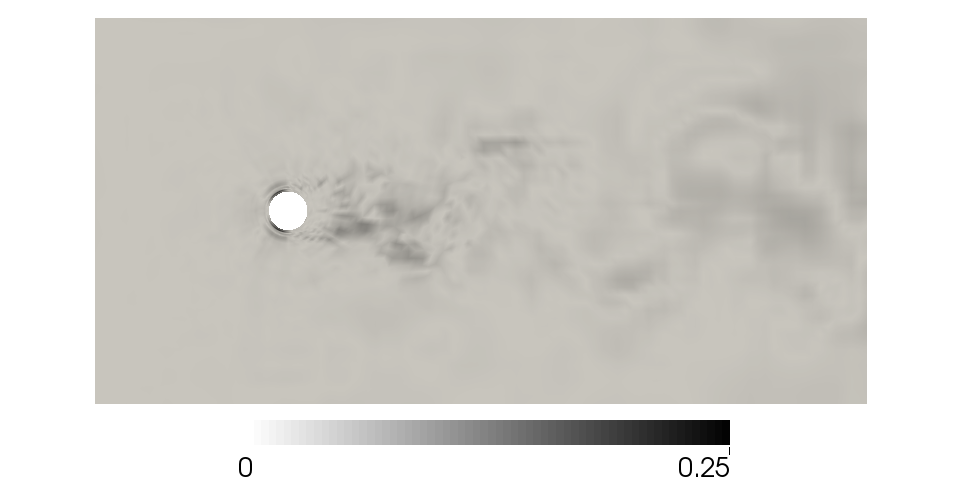}
 }
 \hfill
 \subfloat[Re=250]{%
 \includegraphics[width=0.49\columnwidth]{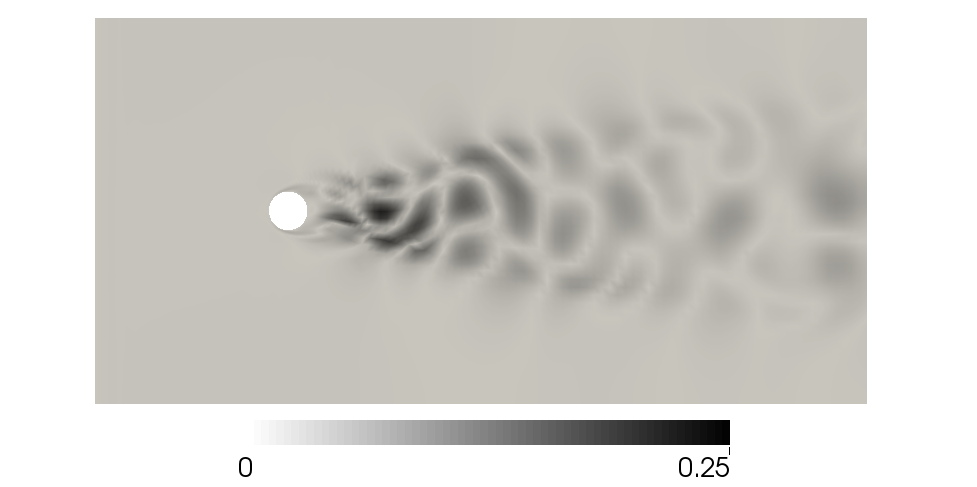}
 \hfill
 \includegraphics[width=0.49\columnwidth]{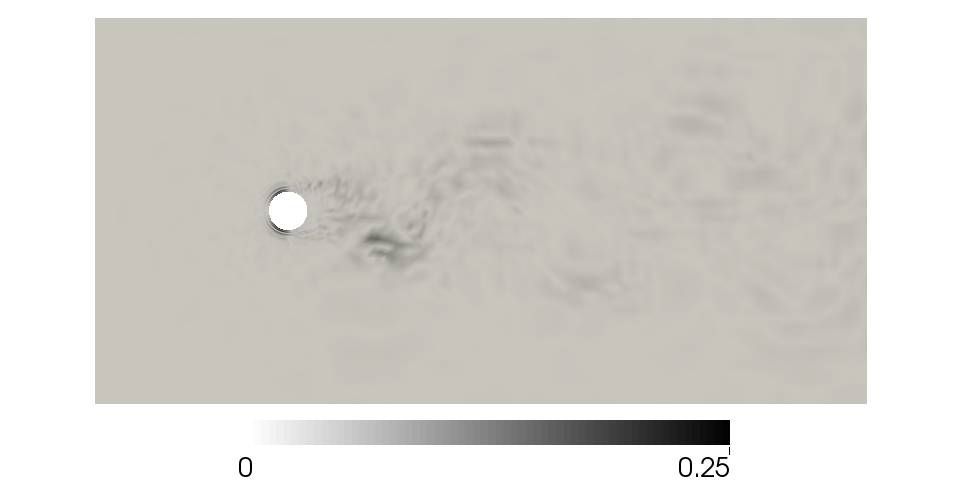}
 }
 \hfill
 \subfloat[Re=350]{%
 \includegraphics[width=0.49\columnwidth]{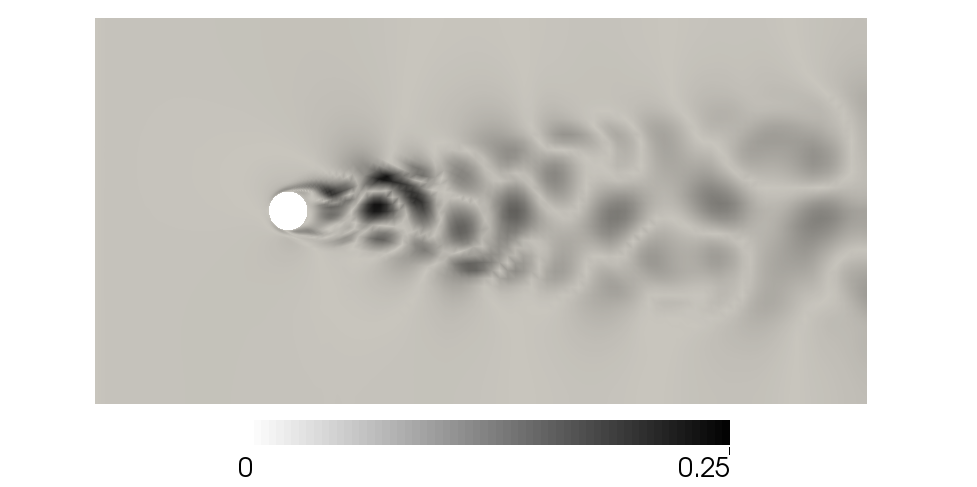}
 \hfill
 \includegraphics[width=0.49\columnwidth]{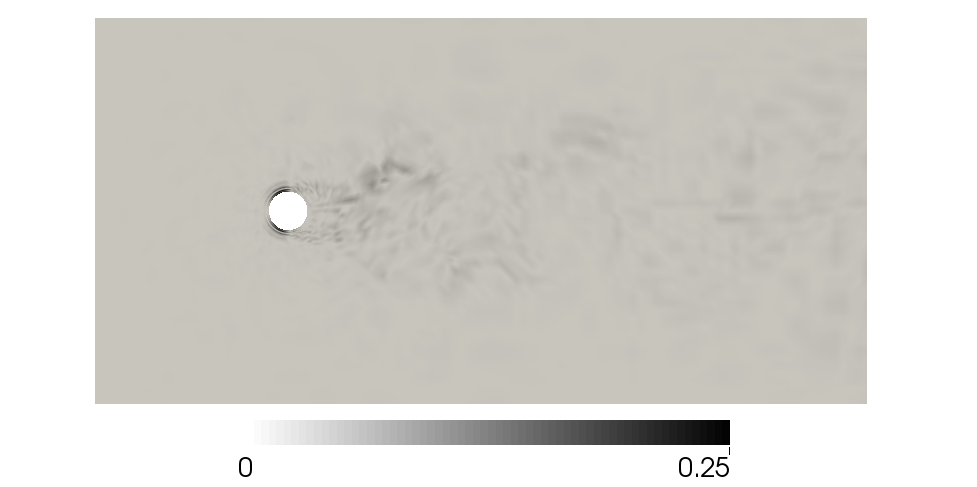}
 }
 \caption{Absolute error in the horizontal velocity reconstruction at the final timestep for the 3 test datasets using POD (left) and our autoencoder (right) with $d_{latent}=4$.}
 \label{fig:ex2_result}
\end{figure} 

A selection of the results are tabulated in \cref{tab:ex2_results}.  Of particular note is that the autoencoder-based reconstructions are consistently more accurate than the optimal linear (POD) reconstruction for small $d_{latent}$.  The complete results are included in \cref{tab:training_flow} in \cref{sec:AppendixA}.  

\begin{table}[h]
  \begin{center}
    \caption{Performance for laminar flow experiment with $d_{latent}=4$.  Reported values correspond to relative $L^2$ error of the velocity reconstruction averaged over snapshots.}
    \label{tab:ex2_results}
    \begin{tabular}{c|c|c|c|c} 
      \textbf{Mixing} & \textbf{Training} & \multicolumn{3}{c}{\textbf{Test Snapshots}}\\
      \textbf{Set} & \textbf{Snapshots} & $Re=150$ & $Re=250$ & $Re=350$  \\
      \hline
      (b) & 6.021e-03 & 1.010e-02 & 9.335e-03 & 8.804e-03 \\
      (e) & 7.580e-03 & 1.082e-02 & 1.078e-02 & 1.039e-02 \\
      (g) & 7.985e-03 & 1.315e-02 & 1.259e-02 & 1.114e-02 \\
      POD & 4.833e-02 & 4.633e-02 & 4.080e-02 & 4.751e-02 \\
    \end{tabular}
  \end{center}
\end{table}  

Relative to the first experiment, the accuracy of the trained autoencoders was insensitive to both of the hyperparameters we varied.  There was not a clear winner amongst the mixing operator sets. Among operator sets with $m > 3$, the loss values are all so close together that slightly different training procedures may lead to different relative rankings, indicating that many of the combinations of mixing operators chosen may be suitable for this transient laminar flow application.  We suspect that this is due to the relative lack of variability between the training and test sets owed to the periodic nature of the flow and the consistent flow direction.

However, operator sets $\mathcal{M}_b$ and $\mathcal{M}_d$ are consistently among the top performers, which indicates that splitting the Laplace operator $\Delta=\Delta_x+\Delta_y$ is likely not worthwhile for this application, possibly due to the invariance of the inlet flow direction across the dataset.  Once again, POD achieves comparable accuracy to the autoencoder with $d_{latent}=16$ although it is still consistently worse than even the worst performing autoencoder at that size (see \cref{fig:ex2_b}).  \cref{fig:ex2_lift} and \cref{fig:ex2_result} show comparisons between the autoencoders with $d_{latent}=4$ and POD.  
\cref{fig:ex2_lift} shows the temporal distributions of the lift force on the cylinder calculated from the reconstructed pressure and the $L^2$ error in the velocity reconstruction, while \cref{fig:ex2_result} shows the spatial distribution of the reconstruction error in the horizontal velocity at the final timestep.
For small values of $d_{latent}$, our autoencoders are more than an order of magnitude more accurate than POD.
Across all values of $d_{latent}$ value is observed in using differential operators over a traditional GCN ($\mathcal{M}_g$).

The fact that increasing $d_{latent}$ did not result in increased accuracy indicates that the system's dynamic response is adequately captured by the lower dimensional space, which is not surprising, given that the state evolves on a 2D manifold parameterized by Reynolds number and time.  However, as seen in the advection-diffusion example, the existance of a low-dimensional solution manifold is no guarantee that our autoencoder will be able to consistently learn its structure.  Although our autoencoders are able to learn a reasonable 2D manifold, there is a considerable accuracy benefit in providing the network with extra flexibility by setting $d_{latent}=4$.

\subsection{Inviscid Supersonic Flow Over a Wedge}

\begin{figure}[h]
 \includegraphics[width=\columnwidth]{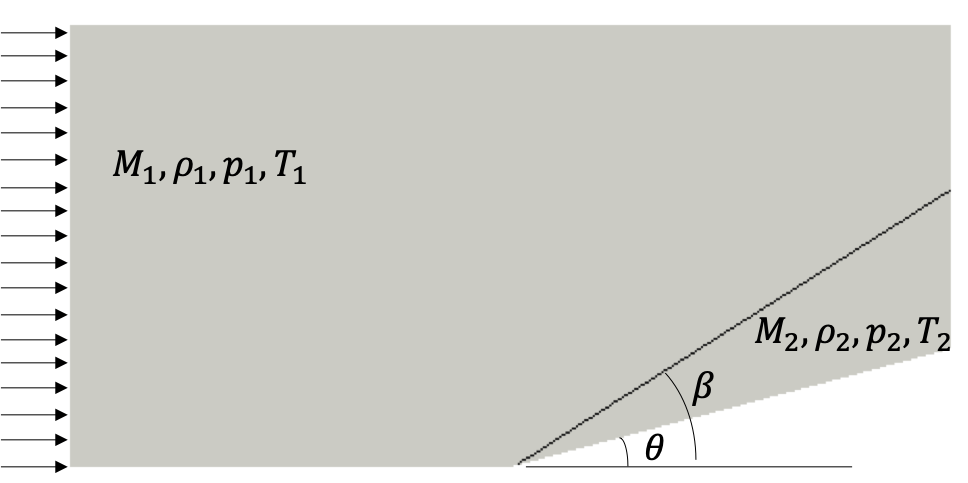}
 \caption{Steady inviscid supersonic flow over a wedge results in an oblique shock originating at the corner.  The flow state is piecewise constant with the inlet and outlet conditions shown.}
 \label{fig:ex7}
\end{figure}

The third and final experiment involves steady, 2D supersonic inviscid flow. The geometry includes a surface with a sharp corner turning into the flow to induce an oblique shock, as shown in \cref{fig:ex7}.  The angle of the wall is $\theta=\ang{15}$.  The governing equations
\begin{equation}
\begin{aligned}
\frac{\partial \rho}{\partial t} + \nabla\cdot\left(\rho \vec{u}\right) &= 0, \\
\frac{\partial \left(\rho\vec{u}\right)}{\partial t} + \nabla\cdot\left(\rho\vec{u}\vec{u}^T\right) + \nabla P &= 0, \text{ and} \\
\frac{\partial(\rho E)}{\partial t} + \nabla\cdot\left(\rho H \vec{u}\right) &= 0
\end{aligned}
\end{equation}
are the inviscid Euler equations (enforcing conservation of mass, momentum, and energy), with $H$ as the total enthalpy and $E$ as the total internal energy.  The equations are closed using the ideal gas law $P=\rho RT/M$.

The inflow boundary conditions are fixed as constant values.  A symmetry boundary condition is prescribed along the bottom surface and open flow boundary conditions are prescribed along the top and right (outflow) surfaces.  The problem is initialized with the inflow conditions and pseudo-timestepping is used to achieve steady-state.  The inflow pressure and temperature are chosen to be $p_1=101352.93$ and $T_1=288.89$, respectively.  The free stream Mach number $M_1$ is varied across training and testing scenarios.  A shock forms at angle $\beta$ to the flow where the wall turns sharply into the flow at the corner.  $\theta$ and $\beta$ are related to $M_1$ by the $\theta-\beta-M$ equation:
\begin{equation}
\label{eq:theta_beta_M}
  \tan\theta = 2\cot\beta\frac{M_1^2\sin^2\beta-1}{M_1^2(\gamma+\cos2\beta)+2}.
\end{equation}
Once $\beta$ is computed from \cref{eq:theta_beta_M}, the downstream conditions can be computed from the following equations:
\begin{equation}
\begin{aligned}
\frac{p_2}{p_1} &= 1 + \frac{2\gamma}{\gamma+1}(M_1^2\sin^2\beta-1), \\
 \frac{\rho_2}{\rho_1} &= \frac{(\gamma+1)M_1^2\sin^2\beta}{(\gamma-1)M_1^2\sin^2\beta+2}, \\
 \frac{T_2}{T_1} &= \frac{p_2}{p_1}\frac{\rho_1}{\rho_2}, \text{ and} \\
 M_2 &= \frac{1}{\sin(\beta-\theta)}\left(\frac{1+\frac{\gamma-1}{2}M_1^2\sin^2\beta}
{\gamma M_1^2\sin^2\beta-\frac{\gamma-1}{2}}\right)^{1/2}.
\end{aligned}
\end{equation}
The conditions ahead of and behind the shock are then given by $\{M_1,\rho_1,p_1,T_1\}$ and $\{M_2,\rho_2,p_2,T_2\}$ respectively.  The local sound speed is given by $c = \sqrt{\gamma\,p/\rho}$.  From the sound speed, the air speed $V = M c = M \sqrt{\gamma\,p/\rho}$ can be computed.  Therefore, the velocity field is ${\bf v} = (u_1,0)$ ahead of the shock and ${\bf v} = (u_2  \, \cos(\theta),u_2  \, \sin(\theta))$ behind the shock, where $u_i = M_i \sqrt{\gamma\,p_i/\rho_i}$ and $i=1,2$.  

The domain is discretized using 12800 quad elements and training data is generated for $M_1\in\lbrace 2,2.1,2.2, \cdots ,5.9, 6.0\rbrace$.  Separate test data is generated for $M_1\in\lbrace 2.25, 3.25, 4.25, 5.25\rbrace$.    

\begin{figure}[h]
 \subfloat[$M_1=2.25$]{%
 \includegraphics[width=0.49\columnwidth]{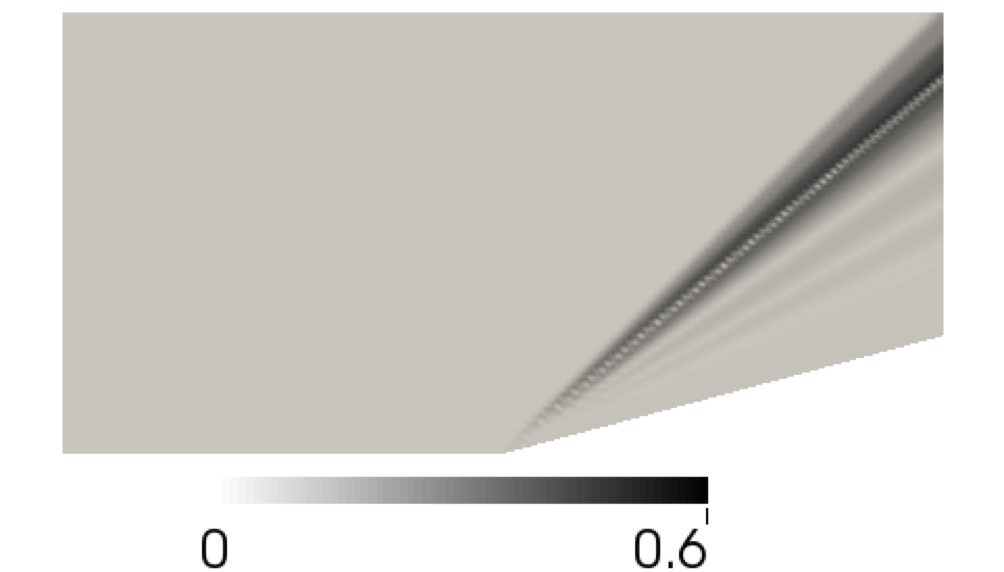}
 \hfill
 \includegraphics[width=0.49\columnwidth]{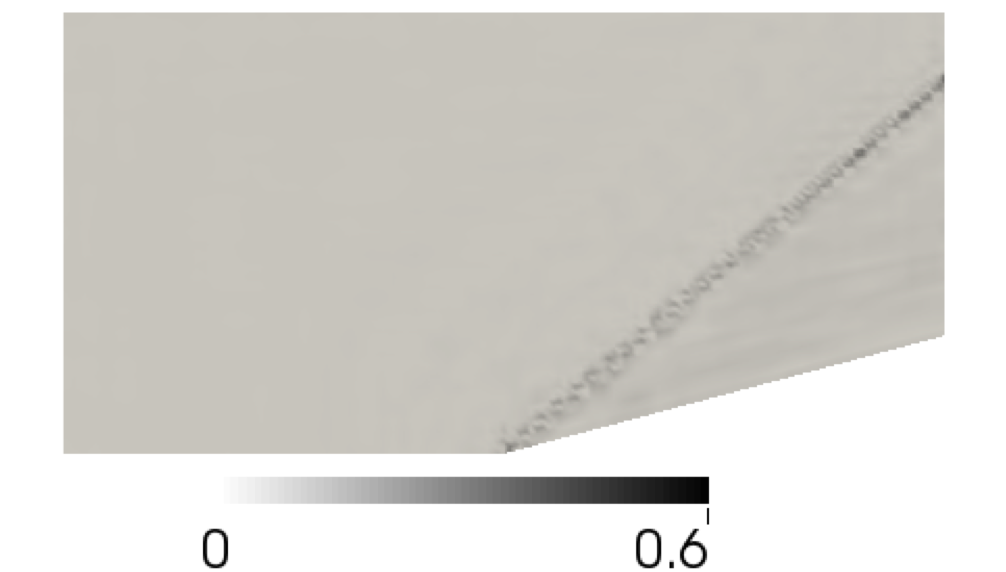}
 }
 \hfill
 \subfloat[$M_1=3.25$]{%
 \includegraphics[width=0.49\columnwidth]{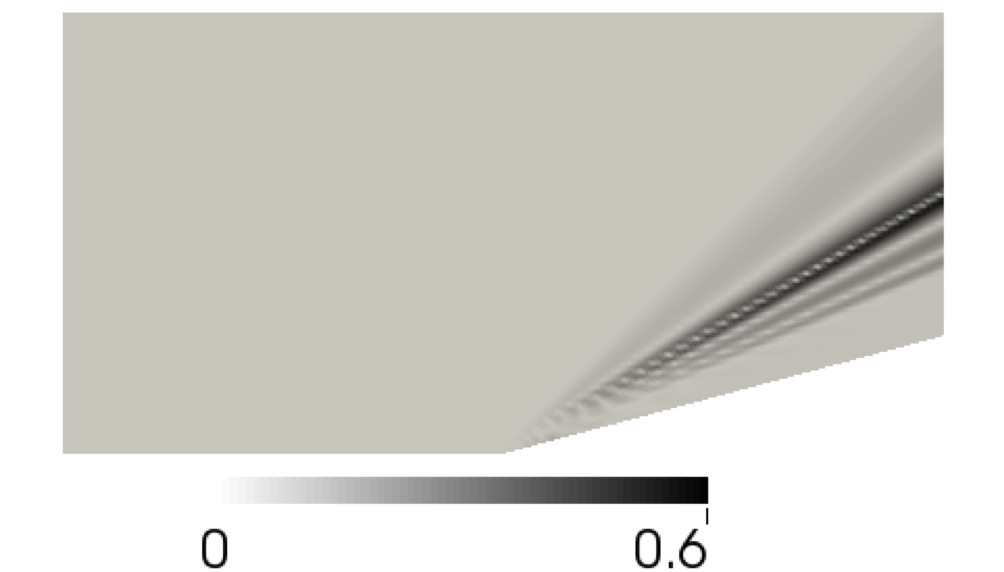}
 \hfill
 \includegraphics[width=0.49\columnwidth]{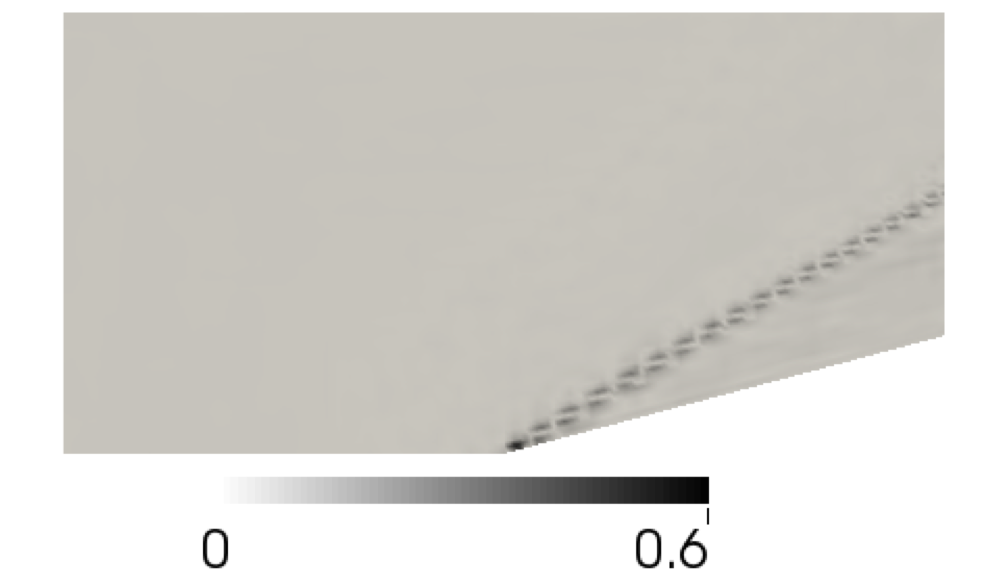}
 }
 \hfill
 \subfloat[$M_1=4.25$]{%
 \includegraphics[width=0.49\columnwidth]{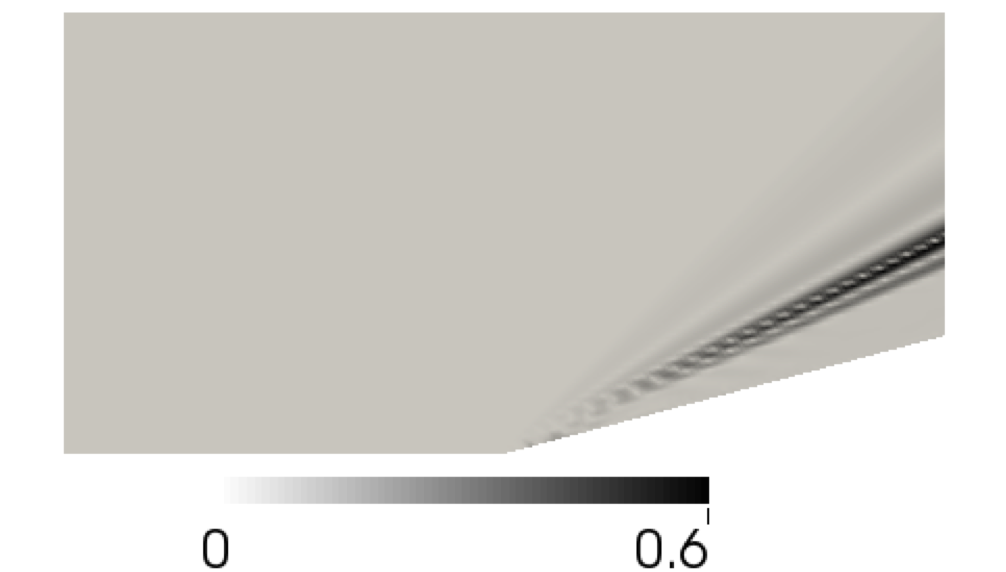}
 \hfill
 \includegraphics[width=0.49\columnwidth]{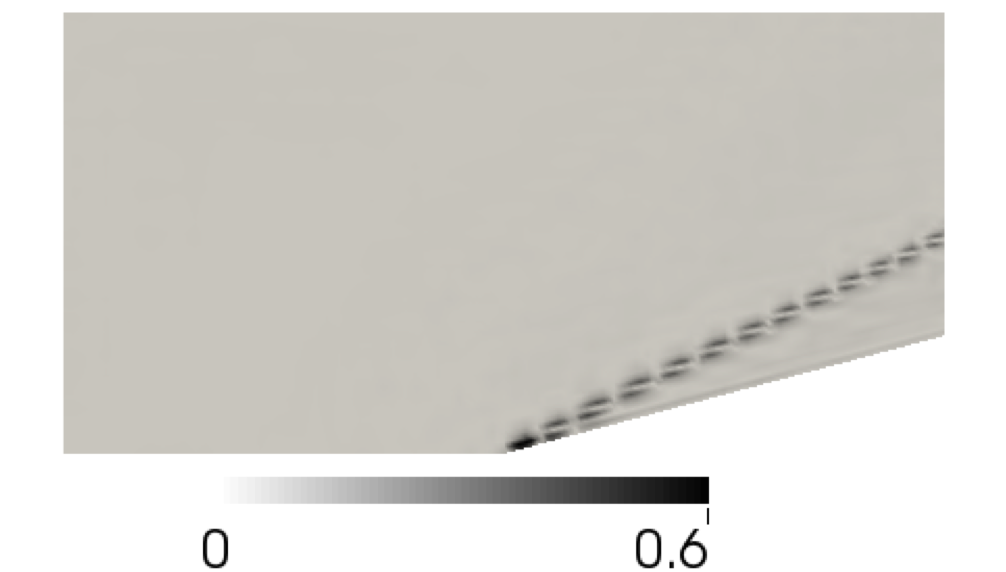}
 }
 \hfill
 \subfloat[$M_1=5.25$]{%
 \includegraphics[width=0.49\columnwidth]{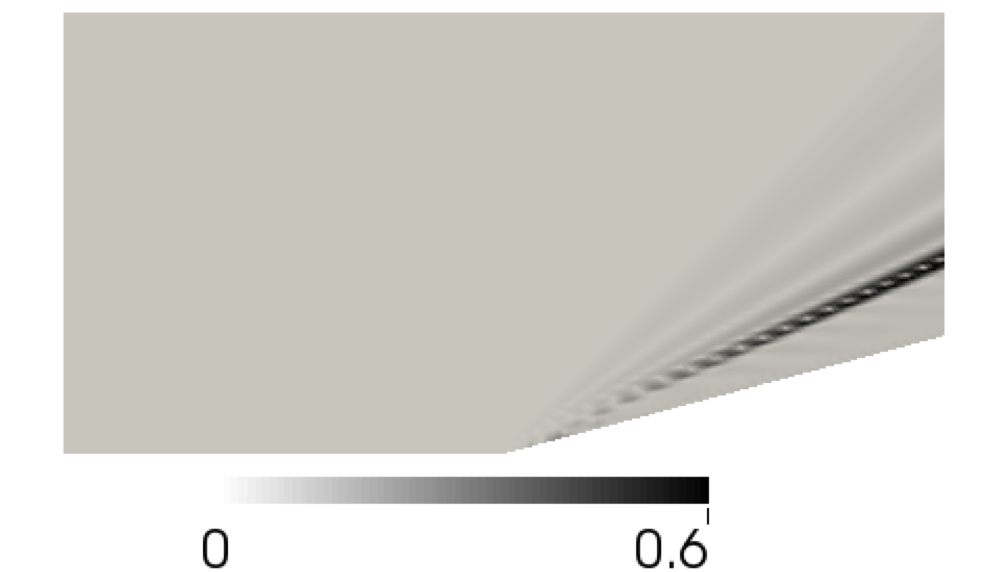}
 \hfill
 \includegraphics[width=0.49\columnwidth]{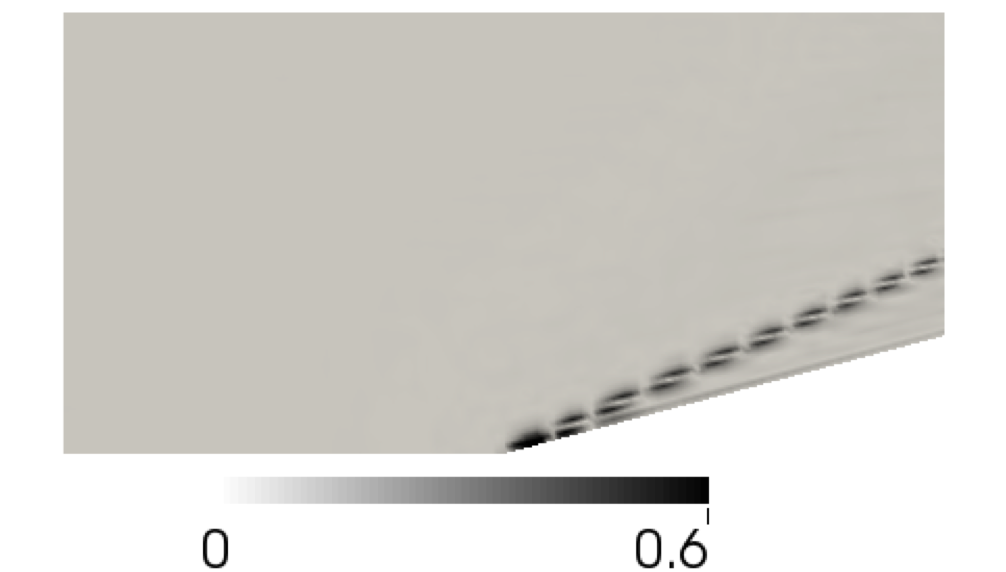}
 }
 \caption{Absolute error in the density reconstruction for the 4 test points using POD (left) and our autoencoder (right) with equivalent latent space dimension ($d_{latent}=4$).}
 \label{fig:ex7_result}
\end{figure}

Full results are tabulated in \cref{tab:training_euler} in \cref{sec:AppendixA}.  The 41 datapoints used for training in this case are considerably fewer than those used for the other experiments, owing to the fact that significantly less information is harvested from a steady-state solution than from a transient solution.  The data could theoretically be augmented by inclusion of pseudo-timestepping data generated during the solution process, but doing so proved unecessary in this case.  However, only having 41 training points restricts the maximum latent dimension of the POD reconstruction to no more than 41.  For this problem, only the density field is reconstructed since the spatial distribution is the same for all solution variables.

Most mixing operator sets performed comparably for this experiment, although we once again observe significant performance degradation for $m<3$.
The autoencoder outperformed the POD reconstruction over the full range of Mach numbers considered for $d_{latent} < 8$.  The POD reconstruction includes large oscilations that pollute the solution both upwind and downwind of the shock.  These oscilations persist (especially for lower Mach numbers) until a relatively large number of modes are retained ($d_{latent}>16$).  In contrast, the autoencoder reconstruction error is localized around the shock front.  For the density reconstruction pictured in \cref{fig:ex7_result}, the amplitude of these errors increases with increasing $M_1$, but only at a rate commensurate with the increase of $\rho_2$.  The autoencoder is able to reconstruct the field data with a very small latent space.  Little performance degradation is seen, even for $d_{latent}=1$, in which case the autoencoder has precisely learned the solution manifold for this problem.

\section{Conclusions}
\label{sec:Conc}

We combined the traditional CNN’s ability to learn physically relevant filters with the GCN’s applicability to handle data defined on unstructured meshes, allowing us to accurately reconstruct the nonlinear solution manifolds of PDEs, which had previously proven challenging for linear manifold methods. Our approach achieves an order of magnitude improvement in accuracy relative to linear methods for the 3 selected exemplar problems. The accuracy is comparable to that achieved by a traditional CNN autoencoder for the transient advection-diffusion problem for which both are applicable. For the other 2 example problems, the traditional CNN is unable to deal with the lack of structure in the spatial discretization, but our proposed approach continues to achieve similar gains in accuracy relative to POD enabling low-cost surrogate models of systems that previously required a much higher dimensional latent space.  Additionally, our networks consistently outperform the traditional GCN which is expressed as a special case of our proposed architecture.  In terms of computational cost, our method is comparable to traditional convolutional autoencoders in terms of FLOPS although further implementation improvements are required to be truly competitive in terms of wall clock time.

The proposed networks represent an extension of CNNs to the processing of data represented using unstructured spatial discretizations. Doing so significantly increases the applicability of CNNs, due to the resulting support for arbitarily complex geometries. The only prerequisite is the ability to compute the required differential and interpolation operators, which should be readily available for any discretization used for the solution of PDEs (including meshless methods). In our implementation, these operators are precomputed, but they could also be computed at run time, which would be useful if training/test data include a variety of spatial discretizations, but would come with additional computational overhead. The extension of the powerful CNN technique to these types of data is important, not only in that it enables low-dimensional nonlinear surrogate models on arbitrarily complex geometries, but also that it enables other tasks requiring low-dimensional solution manifolds of computational data (e.g. synthesizing computational and experimental data, data compression, feature extraction/visualization, anomaly detection) for complex systems.

The reported accuracy improvements are restricted to the datasets considered.  The effects of the various hyperparameters may differ depending on the physics, geometry, solution parameter values, or the training data collected.  The choice of mixing operators is shown to have limited impact on the ultimate accuracy of the method, although highly restrictive sets ($m<3$, which includes the traditional GCN) exhibited markedly worse accuracy.  For the 3 different experiments, different mixing operator sets were shown to be advantageous.  For the advection-diffusion experiment, operator set $\mathcal{M}_e$ was generally best; for the laminar flow experiment, operator set $\mathcal{M}_d$ was generally best; and for the inviscid Euler experiment, operator set $\mathcal{M}_b$ was generally best.  This difference indicates a problem-dependent nature to the optimal choice of mixing operators. 
Additionally, more expressive operator sets (larger $m$) are observed to usually (although not always) result in easier training.  This is consistent with other results in the machine learning literature \cite{buhai2019empirical}.  Different sets of mixing operators (and even different operators entirely) may prove advantageous for other applications.
When approaching a new application, we suggest attempting to use operator set $\mathcal{M}_b$ first and then adding in additional mixing operators if needed.

For the examples considered, the solutions evolve along low-dimensional manifolds, which are able to be successfully represented with $d_{latent}\leq 4$.  Further increasing the latent space dimension  did not generally result in increased accuracy for any of the autoencoder approaches (including the traditional CNN).  Larger latent spaces may be required for different datasets, but it is expected that $d_{latent}$ should not be required to exceed the dimensionality of the underlying solution manifold by much, if at all, which is consistent with prior findings on the use of deep autoencoders for physics applications \cite{lee2018model, lee2019deep}.  A similar network architecture was used for all 3 experiments and minimal effort was made to optimize it.  It is expected that further accuracy improvements are possible if the architecture and training procedure are tailored to a particular dataset.

All of the datasets examined here were 2D.  However, as noted in \cref{sec:Mixing}, the extension to 3D is trivial.  Additionally, the transition from 2D to 3D when using our proposed mixing operator sets is significantly less costly than for traditional CNNs, corresponding to an increase in the number of learnable parameters per convolutional layer of between 25\% and 43\% depending on the mixing operator set relative to the 200\% required for traditional CNNs. 

\FloatBarrier
\appendix
\section{Tables of Loss Values}
\label{sec:AppendixA}
Here, we include detailed summaries of the accuracy achieved in all three experiments, for every value of each hyperparameter considered.  
For the transient-advection diffusion experiment, the solution variable $\alpha\in[0,1]$.  
For the laminar flow experiment, the variables used for training and included in the computation of the loss function are the horizontal velocity $u_x\in(-0.4,1.4)$, the vertical velocity $u_y\in(-0.75,0.75)$, and the pressure $p\in(-0.75,0.6)$.  
For the inviscid Euler experiment, only the density $\rho\in(1.2, 6.2)$ is included since all of the variables share the same spatial distribution.

\begin{table}[h]
  \begin{center}
    \caption{Loss performance for transient advection-diffusion experiment. The `loss' refers to the MSE loss of the solution averaged across solution variables and summed across snapshots.  
For convenience, the most accurate set of mixing operators for each test and for each value of $d_{latent}$ is highlighted.  In those cases where POD was more accurate than the autoencoder, those results are also highlighted. }
    \label{tab:training_advdiff}
    \begin{tabular}{c|c|c|c|c|c|c} 
      & \textbf{Mixing} & \textbf{Training} & \multicolumn{4}{c}{\textbf{Test Loss}}\\
      & \textbf{Set} & \textbf{Loss} & $\theta=\ang{45}$ & $\theta=\ang{135}$ & $\theta=\ang{225}$  & $\theta=\ang{315}$ \\
       \hline
      \multirow{5}{*}{\rotatebox[origin=c]{90}{\parbox[c]{2cm}{\centering $d_{latent}=2$}}}
      & (a) & \cellcolor[gray]{0.9}1.246e-5 & 7.828e-6 & \cellcolor[gray]{0.9}3.759e-6 & 1.868e-5 & 1.036e-3   \\ 
      & (b) & 9.524e-5 & 4.199e-2 & 1.208e-2 & 1.765e-4 & 1.711e-2  \\
      & (c) & 1.596e-5 & \cellcolor[gray]{0.9}5.159e-6 & 1.056e-2 & 6.170e-4 & \cellcolor[gray]{0.9}2.219e-5  \\
      & (d) & 1.261e-5 & 3.217e-2 & 8.677e-6 & \cellcolor[gray]{0.9}4.880e-6 & 1.623e-4  \\ 
      & (e) & 1.335e-5 & 7.714e-6 & 8.158e-6 & 2.840e-3 & 1.048e-2  \\
      \hline
      \multirow{11}{*}{\rotatebox[origin=c]{90}{\parbox[c]{2cm}{\centering $d_{latent}=4$}}}
      & (a) & 4.116e-5 & 7.339e-6 & 8.486e-6 & 1.102e-5 & 7.708e-6 \\
      & (b) & 5.994e-5 & 1.286e-5 & 1.027e-5 & 9.766e-6 & 1.067e-5 \\
      & (c) & 7.141e-5 & 1.906e-5 & 2.149e-5 & 1.269e-5 & 1.577e-5 \\
      & (d) & 4.083e-5 & 1.425e-5 & 1.046e-5 & 7.233e-6 & 2.288e-5 \\
      & (e) & \cellcolor[gray]{0.9}2.273e-5 & \cellcolor[gray]{0.9}7.048e-6 & \cellcolor[gray]{0.9}5.419e-6 & \cellcolor[gray]{0.9}4.553e-6 & \cellcolor[gray]{0.9}4.967e-6 \\
      & (f)  & 2.221e-4 & 3.289e-5 & 3.338e-5 & 3.274e-5 & 3.458e-5 \\
      & (g) & 4.592e-4 & 6.978e-5 & 6.587e-5 & 6.787e-5 & 6.627e-5 \\
      & (h) & 9.388e-1 & 1.345e-1 & 1.345e-1 & 1.345e-1 & 1.345e-1 \\
      & (i)  & 1.645e-3 & 2.442e-4 & 2.352e-4 & 3.084e-4 & 2.285e-4 \\
      &POD & 1.198e-2 & 1.456e-3 & 1.456e-3 & 1.456e-3 & 1.456e-3 \\
      &CNN& 2.770e-5 & 3.523e-5 & 3.456e-5 & 3.599e-5 & 3.608e-5 \\
      \hline
      \multirow{11}{*}{\rotatebox[origin=c]{90}{\parbox[c]{2cm}{\centering $d_{latent}=16$}}}
      & (a) & 2.305e-4 & 3.982e-5 & 4.294e-5 & 3.205e-5 & 4.031e-5 \\
      & (b) & 7.178e-5 & 1.164e-5 & 1.169e-5 & 1.144e-5 & 1.181e-5 \\
      & (c) & \cellcolor[gray]{0.9}9.102e-6 & \cellcolor[gray]{0.9}2.828e-6 & 9.565e-6 & \cellcolor[gray]{0.9}4.527e-6 & \cellcolor[gray]{0.9}3.649e-6 \\
      & (d) & 4.203e-5 & 1.194e-5 & 1.057e-5 & 1.095e-5 & 1.050e-5 \\
      & (e) & 9.437e-6 & 1.005e-5 & \cellcolor[gray]{0.9}7.028e-6 & 5.801e-6 & 9.888e-6 \\
      & (f)  & 9.388e-1 & 1.345e-1 & 1.345e-1 & 1.345e-1 & 1.345e-1 \\
      & (g) & 5.791e-1 & 8.338e-2 & 8.239e-2 & 8.353e-2 & 8.181e-2 \\
      & (h) & 9.388e-1 & 1.345e-1 & 1.345e-1 & 1.345e-1 & 1.345e-1 \\
      & (i)  & 8.142e-5 & 1.378e-5 & 1.781e-5 & 2.017e-5 & 1.397e-5 \\
      &POD & 7.096e-5 & 6.973e-6 & \cellcolor[gray]{0.9}6.973e-6 & 6.973e-6 & 6.973e-6 \\
      &CNN& 3.817e-5 & 4.797e-5 & 4.589e-5 & 5.064e-5 & 4.728e-5 \\
      \hline
      \multirow{11}{*}{\rotatebox[origin=c]{90}{\parbox[c]{2cm}{\centering $d_{latent}=64$}}}
      & (a) & \cellcolor[gray]{0.9}1.203e-5 & 4.078e-6 & 7.811e-6 & 5.454e-6 & 6.364e-6 \\
      & (b) & 7.620e-5 & 1.188e-5 & 1.505e-5 & 1.392e-5 & 1.410e-5 \\
      & (c) & 2.376e-5 & 3.756e-6 & 1.024e-5 & 9.465e-6 & 3.630e-5 \\
      & (d) & 2.064e-4 & 3.063e-5 & 3.011e-5 & 3.036e-5 & 3.010e-5 \\
      & (e) & 1.995e-5 & \cellcolor[gray]{0.9}3.676e-6 & \cellcolor[gray]{0.9}4.691e-6 & \cellcolor[gray]{0.9}4.461e-6 & \cellcolor[gray]{0.9}5.268e-6 \\
      & (f)  & 7.820e-2 & 1.120e-2 & 1.122e-2 & 1.120e-2 & 1.119e-2  \\
      & (g) & 1.663e-4 & 2.549e-5 & 2.357e-5 & 3.691e-5 & 2.415e-5 \\
      & (h) & 9.388e-2 & 1.344e-1 & 1.345e-1 & 1.345e-1 & 1.345e-1 \\
      & (i)  & 4.366e-2 & 1.824e-3 & 1.003e-2 & 1.623e-3 & 1.122e-2 \\
      &POD & \cellcolor[gray]{0.9}1.002e-9 & \cellcolor[gray]{0.9}1.277e-10 & \cellcolor[gray]{0.9}1.274e-10 & \cellcolor[gray]{0.9}1.274e-10 & \cellcolor[gray]{0.9}1.275e-10 \\
      &CNN& 2.872e-5 & 3.978e-5 & 3.923e-5 & 3.937e-5 & 3.843e-5 \\
    \end{tabular}
  \end{center}
\end{table}

\begin{table}[h]
  \begin{center}
    \caption{$L^2$ norm performance for laminar flow experiment.  Reported value are the relative $L^2$ error norm of the reconstructed velocity.  All results converged in fewer than 2,000 epochs.}
    \label{tab:training_flow}
    \begin{tabular}{c|c|c|c|c|c} 
      & \textbf{Mixing} & \textbf{Training} & \multicolumn{3}{c}{\textbf{Test Loss}}\\
      & \textbf{Set} & \textbf{Loss} & $Re=150$ & $Re=250$ & $Re=350$  \\
      \hline
      \multirow{10}{*}{\rotatebox[origin=c]{90}{\parbox[c]{2cm}{\centering $d_{latent}=2$}}}
      & (a) & 1.471e-02 & 4.783e-02 & 2.486e-02 & 1.994e-02 \\
      & (b) & 1.304e-02 & 5.602e-02 & 2.511e-02 & 1.825e-02 \\
      & (c) & 1.509e-02 & 5.565e-02 & 2.571e-02 & 1.936e-02 \\
      & (d) & 2.439e-02 & 3.677e-02 & 3.452e-02 & 2.583e-02 \\
      & (e) & 2.100e-02 & 3.764e-02 & 2.634e-02 & 2.240e-02 \\
      & (f) & 2.111e-02 & 3.879e-02 & 2.756e-02 & 2.317e-02 \\
      & (g) & 2.163e-02 & 3.674e-02 & 2.883e-02 & 2.483e-02 \\
      & (h) & 1.144e-01 & 1.339e-01 & 1.444e-01 & 1.502e-01 \\
      & (i) & 1.568e-02 & 4.924e-02 & 2.423e-02 & 2.341e-02 \\
      & (POD) & 8.933e-02 & 7.426e-02 & 6.928e-02 & 7.513e-02 \\
      \hline
      \multirow{10}{*}{\rotatebox[origin=c]{90}{\parbox[c]{2cm}{\centering $d_{latent}=4$}}}
      & (a) & 6.644e-03 & 1.237e-02 & 1.202e-02 & 1.008e-02 \\
      & (b) & 6.021e-03 & 1.010e-02 & 9.335e-03 & 8.804e-03 \\
      & (c) & 6.476e-03 & 1.337e-02 & 1.109e-02 & 1.021e-02 \\
      & (d) & 6.481e-03 & 1.165e-02 & 1.025e-02 & 9.434e-03 \\
      & (e) & 7.580e-03 & 1.082e-02 & 1.078e-02 & 1.039e-02 \\
      & (f) & 8.170e-03 & 1.322e-02 & 1.214e-02 & 1.120e-02 \\
      & (g) & 7.985e-03 & 1.315e-02 & 1.259e-02 & 1.114e-02 \\
      & (h) & 2.339e-02 & 2.913e-02 & 2.633e-02 & 2.551e-02 \\
      & (i) & 8.643e-03 & 1.322e-02 & 1.223e-02 & 1.111e-02 \\
      & (POD) & 4.833e-02 & 4.633e-02 & 4.080e-02 & 4.751e-02 \\
      \hline
      \multirow{10}{*}{\rotatebox[origin=c]{90}{\parbox[c]{2cm}{\centering $d_{latent}=8$}}}
      & (a) & 5.964e-03 & 1.039e-02 & 9.639e-03 & 9.110e-03 \\
      & (b) & 5.805e-03 & 9.268e-03 & 8.879e-03 & 8.399e-03 \\
      & (c) & 5.536e-03 & 9.206e-03 & 9.070e-03 & 8.606e-03 \\
      & (d) & 8.607e-03 & 1.223e-02 & 1.187e-02 & 1.167e-02 \\
      & (e) & 9.566e-03 & 1.236e-02 & 1.233e-02 & 1.242e-02 \\
      & (f) & 7.375e-03 & 1.192e-02 & 1.126e-02 & 1.036e-02 \\
      & (g) & 7.186e-03 & 1.205e-02 & 1.170e-02 & 1.043e-02 \\
      & (h) & 2.084e-02 & 2.551e-02 & 2.430e-02 & 2.402e-02 \\
      & (i) & 7.356e-03 & 1.162e-02 & 1.096e-02 & 1.032e-02 \\
      & (POD) & 2.742e-02 & 3.224e-02 & 3.446e-02 & 3.688e-02 \\
      \hline
      \multirow{10}{*}{\rotatebox[origin=c]{90}{\parbox[c]{2cm}{\centering $d_{latent}=16$}}}
      & (a) & 5.871e-03 & 1.061e-02 & 9.553e-03 & 8.961e-03 \\
      & (b) & 6.314e-03 & 1.063e-02 & 9.855e-03 & 9.084e-03 \\
      & (c) & 5.315e-03 & 8.677e-03 & 9.069e-03 & 8.498e-03 \\
      & (d) & 8.730e-03 & 1.232e-02 & 1.154e-02 & 1.130e-02 \\
      & (e) & 8.388e-03 & 1.167e-02 & 1.137e-02 & 1.123e-02 \\
      & (f) & 7.362e-03 & 1.199e-02 & 1.143e-02 & 1.071e-02 \\
      & (g) & 7.561e-03 & 1.190e-02 & 1.125e-02 & 1.087e-02 \\
      & (h) & 1.142e-01 & 1.338e-01 & 1.444e-01 & 1.502e-01 \\
      & (i) & 7.361e-03 & 1.097e-02 & 1.035e-02 & 1.011e-02 \\
      & (POD) & 1.889e-02 & 2.021e-02 & 1.926e-02 & 2.075e-02 \\
    \end{tabular}
  \end{center}
\end{table}    

\begin{table}[h]
  \begin{center}
    \caption{$L^2$ performance for Euler experiment.  Reported values are the relative $L^2$ error norm of the reconstructed density.  All results converged in fewer than 10,000 epochs.}
    \label{tab:training_euler}
    \begin{tabular}{c|c|c|c|c|c|c} 
      & \textbf{Mixing} & \textbf{Training} & \multicolumn{4}{c}{\textbf{Test Loss}}\\
      & \textbf{Set} & \textbf{Loss} & $M_1=2.25$ & $M_1=3.25$ & $M_1=4.25$ & $M_1=5.25$ \\
      \hline
      \multirow{10}{*}{\rotatebox[origin=c]{90}{\parbox[c]{2cm}{\centering $d_{latent}=1$}}}
& (a) & 1.593e-02 & 4.002e-03 & 5.916e-03 & 9.158e-03 & 1.281e-02 \\
& (b) & 1.551e-02 & 3.483e-03 & 5.739e-03 & 9.087e-03 & 1.279e-02 \\
& (c) & 1.560e-02 & 3.579e-03 & 5.762e-03 & 9.107e-03 & 1.278e-02 \\
& (d) & 1.587e-02 & 3.810e-03 & 5.896e-03 & 9.171e-03 & 1.284e-02 \\
& (e) & 1.841e-02 & 5.029e-03 & 6.970e-03 & 1.060e-02 & 1.441e-02 \\
& (f) & 1.961e-02 & 6.103e-03 & 7.166e-03 & 1.059e-02 & 1.489e-02 \\
& (g) & 1.231e-01 & 8.232e-02 & 5.029e-02 & 4.711e-02 & 7.018e-02 \\
& (h) & 1.230e-01 & 8.227e-02 & 5.027e-02 & 4.704e-02 & 7.021e-02 \\
& (i) & 1.588e-02 & 3.654e-03 & 5.975e-03 & 9.300e-03 & 1.299e-02 \\
& POD & 7.352e-02 & 3.872e-02 & 4.478e-02 & 3.395e-02 & 2.817e-02 \\
      \hline
      \multirow{10}{*}{\rotatebox[origin=c]{90}{\parbox[c]{2cm}{\centering $d_{latent}=2$}}}
& (a) & 1.644e-02 & 4.344e-03 & 6.165e-03 & 9.344e-03 & 1.291e-02 \\
& (b) & 1.549e-02 & 3.443e-03 & 5.746e-03 & 9.077e-03 & 1.277e-02 \\
& (c) & 1.585e-02 & 3.846e-03 & 5.876e-03 & 9.185e-03 & 1.283e-02 \\
& (d) & 1.686e-02 & 4.917e-03 & 6.255e-03 & 9.402e-03 & 1.299e-02 \\
& (e) & 1.566e-02 & 3.566e-03 & 5.777e-03 & 9.136e-03 & 1.281e-02 \\
& (f) & 1.912e-02 & 6.923e-03 & 7.482e-03 & 1.023e-02 & 1.369e-02 \\
& (g) & 1.230e-01 & 8.229e-02 & 5.033e-02 & 4.700e-02 & 7.015e-02 \\
& (h) & 2.190e-02 & 1.160e-02 & 9.602e-03 & 1.069e-02 & 1.353e-02 \\
& (i) & 1.599e-02 & 3.875e-03 & 5.972e-03 & 9.255e-03 & 1.297e-02 \\
& POD & 5.281e-02 & 3.093e-02 & 2.685e-02 & 2.867e-02 & 1.395e-02 \\
      \hline
      \multirow{10}{*}{\rotatebox[origin=c]{90}{\parbox[c]{2cm}{\centering $d_{latent}=4$}}}
& (a) & 1.594e-02 & 4.030e-03 & 5.850e-03 & 9.160e-03 & 1.282e-02 \\
& (b) & 1.551e-02 & 3.461e-03 & 5.754e-03 & 9.081e-03 & 1.278e-02 \\
& (c) & 1.562e-02 & 3.567e-03 & 5.783e-03 & 9.120e-03 & 1.282e-02 \\
& (d) & 1.607e-02 & 4.085e-03 & 5.953e-03 & 9.213e-03 & 1.288e-02 \\
& (e) & 1.603e-02 & 3.792e-03 & 5.935e-03 & 9.202e-03 & 1.292e-02 \\
& (f) & 1.578e-02 & 3.503e-03 & 5.856e-03 & 9.253e-03 & 1.305e-02 \\
& (g) & 1.613e-02 & 3.757e-03 & 5.891e-03 & 9.358e-03 & 1.327e-02 \\
& (h) & 1.230e-01 & 8.236e-02 & 5.025e-02 & 4.699e-02 & 7.014e-02 \\
& (i) & 1.561e-02 & 3.502e-03 & 5.811e-03 & 9.123e-03 & 1.283e-02 \\
& POD & 3.297e-02 & 2.593e-02 & 1.698e-02 & 1.257e-02 & 8.830e-03 \\
      \hline
      \multirow{10}{*}{\rotatebox[origin=c]{90}{\parbox[c]{2cm}{\centering $d_{latent}=8$}}}
& (a) & 1.562e-02 & 3.605e-03 & 5.773e-03 & 9.097e-03 & 1.279e-02 \\
& (b) & 1.546e-02 & 3.369e-03 & 5.730e-03 & 9.087e-03 & 1.279e-02 \\
& (c) & 1.561e-02 & 3.589e-03 & 5.765e-03 & 9.110e-03 & 1.280e-02 \\
& (d) & 1.641e-02 & 4.688e-03 & 6.022e-03 & 9.201e-03 & 1.287e-02 \\
& (e) & 1.565e-02 & 3.577e-03 & 5.809e-03 & 9.106e-03 & 1.285e-02 \\
& (f) & 1.567e-02 & 3.555e-03 & 5.864e-03 & 9.153e-03 & 1.289e-02 \\
& (g) & 1.647e-02 & 3.970e-03 & 6.119e-03 & 9.511e-03 & 1.345e-02 \\
& (h) & 1.230e-01 & 8.229e-02 & 5.022e-02 & 4.704e-02 & 7.021e-02 \\
& (i) & 1.568e-02 & 3.604e-03 & 5.844e-03 & 9.152e-03 & 1.284e-02 \\
& POD & 1.690e-02 & 1.389e-02 & 8.644e-03 & 5.928e-03 & 6.802e-03 \\
    \end{tabular}
  \end{center}
\end{table}

\FloatBarrier

\section{Network Architecture}
\label{sec:AppendixB}

\tikzstyle{ann} = [draw=none,fill=none,right]
\tikzstyle{block} = [rectangle, draw, fill=white, minimum width=0.1cm, minimum height=.1cm, node distance=2cm]
\tikzstyle{line} = [draw, -latex']

\begin{tikzpicture}[auto]
    \node [block, minimum width=10cm, minimum height=0.1cm] (input) {Input};
    \node [ann, left of=input, node distance=5cm+1em] (input_dimx) {$C$};
    \node [ann, above of=input, node distance=1.5em] (input_dimy) {$N_0$};
    \node [ann, below of=input, node distance=1.5em] (conv_op1) {Encoder Convolutional Layer};

    \node [block, minimum width=7cm, minimum height=1cm, below of=conv_op1, node distance=1.6cm] (hidden1) {};
    \node [ann, left of=hidden1, node distance=3.5cm+1em] (hidden1_dimx) {$32$};
    \node [ann, above of=hidden1, node distance=2.1em] (hidden1_dimy) {$N_1$};
    \node [ann, below of=hidden1, node distance=2.1em] (conv_op2) {Encoder Convolutional Layer};
    \node [ann, below of=conv_op2, node distance=2.1em, rotate=90] (dots) {$\cdots$};
    
    \node [block, minimum width=3cm, minimum height=1cm, below of=dots, node distance=1.5cm] (hidden3) {};
    \node [ann, left of=hidden3, node distance=1.5cm+1em] (hidden3_dimx) {$32$};
    \node [ann, above of=hidden3, node distance=2.1em] (hidden3_dimy) {$N_d$};
    \node [ann, below of=hidden3, node distance=2.1em] (fc_op1) {Dense Network};
           
    \node [block, minimum width=1cm, minimum height=0.1cm, below of=fc_op1, node distance=1.2cm] (fc2) {};
    \node [ann, left of=fc2, node distance=0.5cm+1em] (fc2_dimx) {$1$};
    \node [ann, above of=fc2, node distance=1em] (fc2_dimy) {$d_{latent}$};
    \node [ann, below of=fc2, node distance=1em] (fc_op3) {Dense Network};
    
    \node [block, minimum width=3cm, minimum height=1cm, below of=fc_op3, node distance=1.7cm] (conv1) {};
	\node [ann, left of=conv1, node distance=1.5cm+1em] (conv1_dimx) {$64$};
    \node [ann, above of=conv1, node distance=.5cm+1em] (conv1_dimy) {$N_d$};
    \node [ann, below of=conv1, node distance=.5cm+1em] (conv1_op4) {Decoder Convolutional Layer};
    
        \node [block, minimum width=3cm, minimum height=1cm, below of=conv1_op4, node distance=1.7cm] (conv2) {};
	\node [ann, left of=conv2, node distance=1.5cm+1em] (conv2_dimx) {$64$};
    \node [ann, above of=conv2, node distance=.5cm+1em] (conv2_dimy) {$N_1$};
    \node [ann, below of=conv2, node distance=.5cm+1em] (conv2_op5) {Decoder Convolutional Layer};
    
    \node [ann, below of=conv2_op5, node distance=2.1em, rotate=90] (dots2) {$\cdots$};
    
    \node [block, minimum width=10cm, minimum height=1cm, below of=dots2, node distance=1.5cm] (conv3) {};
	\node [ann, left of=conv3, node distance=5cm+1em] (conv3_dimx) {$64$};
    \node [ann, above of=conv3, node distance=.5cm+1em] (conv3_dimy) {$N_0$};
    \node [ann, below of=conv3, node distance=.5cm+1em] (conv3_op6) {Output Layer};	
    
    \node [block, minimum width=10cm, minimum height=0.1cm, below of=conv3_op6, node distance=1.4cm] (output) {Output};
    \node [ann, left of=output, node distance=5cm+1em] (output_dimx) {$C$};
    \node [ann, above of=output, node distance=1.5em] (output_dimy) {$N_0$};
    
    \path [line] (conv_op1) -- (hidden1_dimy);
	\path [line] (fc_op1) -- (fc2_dimy);
	
	\path [line] (conv1_op4) -- (conv2_dimy);
	\path [line] (fc_op3) -- (conv1_dimy);
    \path [line] (conv3_op6) -- (output_dimy);
	
\end{tikzpicture}

\FloatBarrier
\section*{Acknowledgments}
Supported by the Laboratory Directed Research and Development program at Sandia National Laboratories, a multimission laboratory managed and operated by National Technology and Engineering Solutions of Sandia LLC, a wholly owned subsidiary of Honeywell International Inc. for the U.S. Department of Energy’s National Nuclear Security Administration under contract DE-NA0003525.

\bibliographystyle{siamplain}
\bibliography{references}
\end{document}


\maketitle

\section{A detailed example}

Here we include some equations and theorem-like environments to show
how these are labeled in a supplement and can be referenced from the
main text.
Consider the following equation:
\begin{equation}
  \label{eq:suppa}
  a^2 + b^2 = c^2.
\end{equation}
You can also reference equations such as \cref{eq:matrices,eq:bb} 
from the main article in this supplement.

\lipsum[100-101]

\begin{theorem}
  An example theorem.
\end{theorem}

\lipsum[102]
 
\begin{lemma}
  An example lemma.
\end{lemma}

\lipsum[103-105]

Here is an example citation: \cite{KoMa14}.

\section[Proof of Thm]{Proof of \cref{thm:bigthm}}
\label{sec:proof}

\lipsum[106-112]

\section{Additional experimental results}
\Cref{tab:foo} shows additional
supporting evidence. 

\begin{table}[htbp]
{\footnotesize
  \caption{Example table}  \label{tab:foo}
\begin{center}
  \begin{tabular}{|c|c|c|} \hline
   Species & \bf Mean & \bf Std.~Dev. \\ \hline
    1 & 3.4 & 1.2 \\
    2 & 5.4 & 0.6 \\ \hline
  \end{tabular}
\end{center}
}
\end{table}

\bibliographystyle{siamplain}
\bibliography{references}

%% file: ex_shared.tex
\usepackage{lipsum}
\usepackage{amsfonts}
\usepackage{graphicx}
\usepackage{epstopdf}
\usepackage{algorithmic}
\ifpdf
  \DeclareGraphicsExtensions{.eps,.pdf,.png,.jpg}
\else
  \DeclareGraphicsExtensions{.eps}
\fi

\newsiamremark{remark}{Remark}
\newsiamremark{hypothesis}{Hypothesis}
\crefname{hypothesis}{Hypothesis}{Hypotheses}
\newsiamthm{claim}{Claim}

\headers{Convolutional Networks for Unstructured Data}{J. Tencer and K. Potter}

\title{A Tailored Convolutional Neural Network for Nonlinear Manifold Learning of Computational Physics Data using Unstructured Spatial Discretizations\thanks{Submitted to the editors June 8, 2020.
\funding{This work was funded by the Laboratory Directed Research and Development program at Sandia National Laboratories}}}

\author{John Tencer\thanks{Sandia National Laboratories, Albuquerque, NM 
  (\email{jtencer@sandia.gov}, \email{kmpotte@sandia.gov}).}
\and Kevin Potter\footnotemark[2]}

\usepackage{amsopn}
